\documentclass[%
 reprint,
 amsmath,amssymb,
  aps,
 prb,
]{revtex4-1}

\usepackage{graphicx}
\usepackage{hyperref}

\begin{document}

\preprint{APS/123-QED}

\title{Relaxation dynamics in the alternating XY chain following a quantum quench}

\author{Kaiyuan Cao}%
 \affiliation{%
 Zhejiang Lab, Hangzhou 311100, P. R. China
}%

\author{Yayun Hu}
 \email{hyy@zhejianglab.com}
\affiliation{Zhejiang Lab, Hangzhou 311100, P. R. China}

\author{Peiqing Tong}
 \email{pqtong@njnu.edu.cn}
\affiliation{Department of Physics and Institute of Theoretical Physics, Nanjing Normal University, Nanjing 210023, P. R. China}
\affiliation{Jiangsu Key Laboratory for Numerical Simulation of Large Scale Complex Systems, Nanjing Normal University, Nanjing 210023, P. R. China}

\author{Guangwen Yang}%
 \email{ygw@tsinghua.edu.cn}
\affiliation{Zhejiang Lab, Hangzhou 311100, P. R. China}
\affiliation{Department of Computer Science and Technology, Tsinghua University, Haidian District, Beijing 100084, P. R. China}%

\author{Peng Liu}%
 \email{liupeng@zju.edu.cn}
\affiliation{Zhejiang Lab, Hangzhou 311100, P. R. China}
\affiliation{College of Information Science and Electronic Engineering, Zhejiang University, Hangzhou 310027, P. R. China}%

\date{\today}

\begin{abstract}
    We investigate the relaxation dynamics of the fermion two-point correlation function $C_{mn}(t)=\langle\psi(t)|c_{m}^{\dag}c_{n}|\psi(t)\rangle$ in the XY chain with alternating nearest-neighbor hopping interaction after a quench. We find that the deviation $\delta C_{mn}(t)=C_{mn}(t)-C_{mn}(\infty)$ decays with time following the power law behavior $t^{-\mu}$, where the exponent $\mu$ depends on whether the quench is to the commensurate phase ($\mu=1$) or incommensurate phase ($\mu=\frac{1}{2}$). This decay of $\delta C_{mn}(t)$ arises from the transient behavior of the double-excited quasiparticle occupations and the transitions between different excitation spectra. Furthermore, we find that the steady value $C_{mn}(\infty)$ only involves the average fermion occupation numbers (i.e. the average excited single particle) over the time-evolved state, which are different from the ground state expectation values. We also observe nonanalytic singularities in the steady value $C_{mn}(\infty)$ for the quench to the critical points of the quantum phase transitions (QPTs), suggesting its potential use as a signature of QPTs.
\end{abstract}

\maketitle


\section{Introduction}

Over the past decade, nonequilibrium many-body dynamics have seen rapid development due to the experimental advances in quantum systems, such as ultracold atoms \cite{nature.2002.419.51, nature.2012.8.325, science.2012.337.1318, Oxford.2012, nature.2015.11.124}, trapped ions \cite{Zhang2017551, Jurcevic2017119}, Rydberg atoms \cite{nature.2017.551.7682}, and others \cite{Choi2017543, Guo201911}. One interesting question in the field concerns the concept of phase transition far from equilibrium. Based on different identification characteristics, two types of dynamical phase transitions (DPTs) are extensively explored. One type refers to the nonanalytic behavior of the early-time evolution of the Loschmidt echo (DPT-I) \cite{PhysRevLett.2013.110.135704, Zvyagin201642, RepProPhys.2018.81.054001, PhysRevB.2016.93.085416, Fogarty201719, Yang2019100, Wang2019122, Zhou202123, Benini202123, Cao2020102, Mondal2022106, Kawabata2023108, Sedlmayr2023108}. Different from the symmetry-breaking phase transition in equilibrium, DPT-I lacks local order parameters and therefore has no well-defined dynamical phase. The other type describes the asymptotic late-time steady state of an order parameter (DPT-II) \cite{PhysRevLett.2009.102.130603, AdvPhys.2010.59.1063, NewJPhys.2010.12.055017}. Unlike DPT-I, DPT-II is the generalization of the equilibrium phase transition to the nonequilibrium system, where the dynamical phase is characterized by the relaxation behavior of the average order parameter \cite{PhysRevB.2016.94.214301, JPA2014.47.255301, JSM2016.064003, PhysRevE.2018.98.032110, PhysRevB.2023.107.134201, PhysRevB.2016.93.184301, PhysRevB.2016.94.064302, PhysRevB.2021.103.024301, PRL2022.128.050601, JPA2014.47.175002, PRL2021.126.210602, PRB2022.105.094304}. Notably, these two types of DPTs are found to be closely related in the long-range quantum Ising chain \cite{Zunkovic2018120}.

Recently, a new type of DPTs has been proposed in the periodically driven systems \cite{PhysRevB.2016.94.214301, JPA.2018.51.334002, PRB2020.102.235154, PhysRevB.2022.105.104303}, in which the dynamical phases are identified by the relaxation behavior of the two-point correlation function $C_{mn}(t)=\langle\psi(t)|c_{m}^{\dag}c_{n}|\psi(t)\rangle$. Specifically, the deviation $\delta C_{mn}(t)=C_{mn}(t)-C_{mn}(\infty)$ of $C_{mn}(t)$ from its long-time steady values decays as a power-law behavior $t^{-\mu}$. The transition of scaling exponent $\mu$ is dependent on the system parameters, suggesting it to characterize the dynamical phase \cite{PhysRevB.2016.94.214301}. This type of DPTs has recently been extended to the quantum systems following a quantum quench, including the noninteracting and interacting integrable systems \cite{PRB.2022.105.054301, PhysRevB.2023.107.075138}. In the quantum XY chain \cite{PRB.2022.105.054301}, it is shown that $\delta C_{mn}(t)$ may decay as $\sim t^{-1/2}$, $\sim t^{-2/3}$, and $\sim t^{-3/4}$, depending solely on whether the post-quench Hamiltonian is in the commensurate phase, the incommensurate phase, or the boundary between the two phases (disorder line), respectively. However, further studies show that the initial state can also influence the relaxation behavior of $\delta C_{mn}(t)$ in the extended XY chain with the gapless phase, where a new scaling exponent $\mu=1$ is found \cite{PhysRevB.2023.108.014303, cao2023dynamical}. Therefore, it is an interesting research topic to continue exploring the new phenomena and the factors that may affect the scaling behavior of $\delta C_{mn}(t)$. Moreover, all of these works only focus on the relaxation behavior of $\delta C_{mn}(t)$, while neglecting the underlying physics of the correlation function $C_{mn}(t)$ tending to the steady value $C_{mn}(\infty)$. For instance, \textit{what are the physical factors that cause the decay of deviation? What information can we obtain from the stable value $C_{mn}(\infty)$?} The answers to these questions will promote our understanding of the non-equilibrium dynamics of quantum systems.

\begin{figure}
  \centering
  \includegraphics[width=1.0\linewidth]{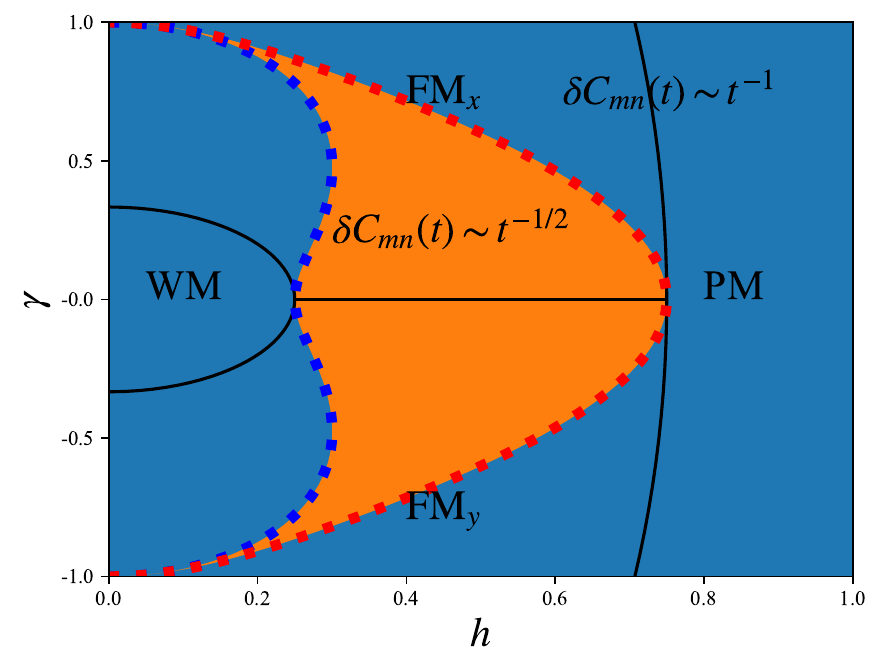}\\
  \caption{Dynamical phase diagram characterized by the power-law decay of $\delta C_{mn}(t)$ in the XY model with the alternating nearest-neighbor hopping interaction $(J_{n}=J_{0}+(-1)^{n}J_{1})$. Here $\alpha=\frac{J_{0}-J_{1}}{J_{0}+J_{1}}=0.5$. The black solid lines denote the critical lines of the QPTs, separating the ferromagnetic (FM$_x$) phase along $x-$direction, the ferromagnetic (FM$_y$) phase along $y-$direction, the paramagnetic (PM) phase, and the phase with weak magnetization (WM). The dotted blue and red lines, known as the disordered lines and described by Eqs.~(\ref{eq:disorder.line.1},\ref{eq:disorder.line.2}), separate the commensurate (the region in black-blue) and incommensurate (the region in orange) phases. }\label{fig:phase.diagram}
\end{figure}

In this paper, we investigate the dynamical relaxation behavior in the XY chain with the alternating nearest-neighbor hopping interaction. There are two reasons for choosing this model as the subject of study. On the one hand, we want to examine the impact of non-uniform structures on the relaxation behavior of $\delta C_{mn}(t)$, since it is well-known that the alternating XY chain exhibits richer quantum phase diagrams and behaviors than those of uniform systems \cite{PhysA.1975.81.319, PhysLettA.1975.53.21, PhysA.1977.89.304, PhysA.1980.100.1, PhysB.2001.304.91, PhysRevB.2009.79.104429, JPhysA.2010.43.505302,PhysRevA.2010.81.060101}. On the other hand, the alternating XY chain can be transformed into a spinless fermion model with two quasiparticle excitation spectra through the Jordan-Wigner transformation. The existence of more than one excitation spectra enables us to gain more physical information about the deviation $\delta C_{mn}(t)$. For instance, we find that the decay of $\delta C_{mn}(t)$ originates from the transients of the average double excited terms and the average quasiparticle transition terms between different excitation spectra. The late one is exactly the multi-band effect, which can not be seen in the uniform XY chain.  Meanwhile, we also find that the alternating structure influences the scaling behavior of $\delta C_{mn}(t)$. As illustrated in Fig.~\ref{fig:phase.diagram}, we discover that $\delta C_{mn}(t)$ follows the power law decay of $\sim t^{-1}$ for the quench to the commensurate phase, which is different from that of $\sim t^{-3/2}$ in the uniform XY model. In addition to the exponents $\mu$ that define the dynamical phase, the steady value $C_{mn}(\infty)$ is found to reflect the behavior of the quantum phase transition (QPT). We find that the steady value $C_{mn}(\infty)$ solely involves the average fermion occupation over the time-evolved states. Of course, these average fermion occupations are different from the ground state expectation value \cite{JstatPhys.2009.135.599, Suzuki2013, PhysRevB.2010.82.144302, PhysRevA.1970.2.1075, PhysRevA.1971.3.786}, however, it is interesting that similar to the ground state expectation value, the steady value $C_{mn}(\infty)$ shows the cusp-like singularities for the quench to the critical points of QPTs, which suggests that $C_{mn}(\infty)$ can be used to characterize the QPT. We also examine the singular behavior of $C_{mn}(\infty)$ in the uniform XY chain to confirm its generality.

The paper is organized as follows. In Sec.~\uppercase\expandafter {\romannumeral2}, we introduce the main properties of the alternating XY chain and give the formula of $C_{mn}(t)$. In Sec.~\uppercase\expandafter {\romannumeral3}, we investigate the dynamical relaxation behavior of $\delta C_{mn}(t)$ through the typical numerical results for quenches to the commensurate phase, incommensurate phase, and disorder lines. In Sec.~\uppercase\expandafter {\romannumeral4}, we study the behavior of the steady value $C_{mn}(\infty)$ and analyze the origin of decay of $\delta C_{mn}(t)$. Finally, in Sec.~\uppercase\expandafter {\romannumeral5} we draw our conclusions.

\section{Models}

We consider the XY chain with alternating nearest-neighbor interactions in the transverse field by the following Hamiltonian\cite{PhysA.1975.81.319, PhysLettA.1975.53.21, PhysA.1977.89.304, PhysA.1980.100.1, PhysB.2001.304.91, PhysRevB.2009.79.104429, JPhysA.2010.43.505302,PhysRevA.2010.81.060101, cao2024quantum}:
\begin{equation}\label{eq:Hamil.XY.alternating}
  \begin{split}
    H & = -\frac{1}{2}\sum_{n=1}^{N}\{\frac{J_{0}+(-1)^{n}J_{1}}{2}[(1+\gamma)\sigma_{n}^{x}\sigma_{n+1}^{x} \\
      & \quad\quad\quad\quad\quad +(1-\gamma)\sigma_{n}^{y}\sigma_{n+1}^{y}]+h\sigma_{n}^{z}\},
  \end{split}
\end{equation}
where $\sigma^{x,y,z}$ are the Pauli matrices, $J_{n}=J_{0}+(-1)^{n}J_{1}$ is the strength of interactions between the nearest-neighbor spins, $\gamma$ is the anisotropic parameter, and $h$ is the external transverse field. The model can be transformed to the XY chain in the alternating field $[h_{n}=h_{0}+(-1)^{n}h_{1}]$ \cite{JPhysCS.2006.51.183, SciRep.2019.9.2871, PhysRevB.2020.102.024425, PhysRevB.2020.101.224304, PhysRevE.2021.104.014134} via the duality transformation \cite{Suzuki2013, PhysRevX.2017.7.031061}.
For convenience, we set the parameter
\begin{equation}
  \alpha = \frac{J_{0}-J_{1}}{J_{0}+J_{1}}
\end{equation}
and $J_{0}+J_{1}=1$ without losing the generality.

The Hamiltonian (\ref{eq:Hamil.XY.alternating}) can be solved by implementing the Jordan-Wigner and Bogoliubov transformations. The diagonal form of the Hamiltonian, describing the spinless free fermion in the double lattice, is given by \cite{PhysB.2001.304.91, JPhysA.2010.43.505302, cao2024quantum}
\begin{equation}\label{eq:diagonal.Hamiltonian}
  H = \sum_{k}[\Lambda_{k1}(\eta_{k1}^{\dag}\eta_{k1}-\frac{1}{2})+\Lambda_{k2}(\eta_{k2}^{\dag}\eta_{k_{2}}-\frac{1}{2})].
\end{equation}
Unlike the homogeneous XY chain, the Hamiltonian (\ref{eq:diagonal.Hamiltonian}) has two quasiparticle excitation spectra, which can be obtained by
\begin{equation}
  \Lambda_{k1}^{2} = \frac{1}{2}(A - \sqrt{B}), \quad \Lambda_{k2}^{2} = \frac{1}{2}(A + \sqrt{B})
\end{equation}
with
\begin{eqnarray}
  A &=& \alpha (1-\gamma^{2})\cos{k}+\frac{1}{2}(1+\gamma^{2})(1+\alpha^{2})+2h^{2}, \\
  B &=& \gamma^{2}(\alpha^{2}-1)^{2}+4h^{2}[1+\alpha^{2}+2\alpha\cos{k}].
\end{eqnarray}
The zero-point (ground state) energy is given by
\begin{equation}
  E_{0} = \sum_{k>0}E_{0k} = -\sum_{k>0}(\Lambda_{k1}+\Lambda_{k2}),
\end{equation}
and the ground state can be expressed as $|GS\rangle = \bigotimes_{k>0}|\text{GS}_{k}\rangle$ with $|\text{GS}_{k}\rangle=|0_{k1}0_{-k1}0_{k2}0_{-k2}\rangle$ for every wave vector $k\in(0,\pi]$.

The critical point of the QPT appears where the band gap closes, corresponding to $\min{(\Lambda_{k1})}=0$. Hence, the critical lines of the phase diagram are given by
\begin{eqnarray}
  h &=& \frac{\sqrt{(1+\alpha)^{2}-\gamma^{2}(1-\alpha)^{2}}}{2}, \gamma\in[-1,1], \\
  h &=& \frac{\sqrt{(1-\alpha)^{2}-\gamma^{2}(1+\alpha)^{2}}}{2}, \gamma\in[-\frac{1-\alpha}{1+\alpha}, \frac{1-\alpha}{1+\alpha}].
\end{eqnarray}
As seen in Fig.~\ref{fig:phase.diagram}, the ground state phase diagram of our model consists of four regions: the ferromagnetic phase along $x-$ direction (FM$_x$), the ferromagnetic phase along $y-$ direction (FM$_y$), the paramagnetic phase (PM), and the phase with weak magnetization (WM). In the WM phase, the spins within a supercell generate a cluster with a small total spin, but between the nearest-neighbor supercells are distributed randomly \cite{cao2024quantum}.

The commensurate and incommensurate phases of the system is determined by whether the minimum of the quasiparticle excitation spectrum is located at the boundary or center of the Brillouin zone. From $ (\Lambda_{k1}\neq0)$
\begin{equation}
  \frac{d\Lambda_{k1}^{2}}{dk} = 2\Lambda_{k1}\frac{d\Lambda_{k1}}{dk}=\Lambda_{k1}\frac{d(A-\sqrt{B})}{dk}=0,
\end{equation}
we obtain that the extreme points are $k=0,\pi$ or the wave vector satisfying
\begin{equation}\label{eq:extreme_point}
  \cos{k} = \frac{2h^{2}}{\alpha(1-\gamma^{2})^{2}}-\frac{\gamma^{2}(\alpha^{2}-1)^{2}}{8\alpha h^{2}}-\frac{1+\alpha^{2}}{2\alpha}.
\end{equation}
By considering the solution of Eq.~(\ref{eq:extreme_point}), we obtain the boundary lines of the commensurate and incommensurate phases given by
\begin{eqnarray}
  h_{d1} &=& \sqrt{\frac{C_{-}}{8}[C_{-}+\sqrt{C_{-}^{2}+4\gamma^{2}(1+\alpha)^{2}}]}, \label{eq:disorder.line.1} \\
  h_{d2} &=& \sqrt{\frac{C_{+}}{8}[C_{+}+\sqrt{C_{+}^{2}+4\gamma^{2}(1-\alpha)^{2}}]}, \label{eq:disorder.line.2}
\end{eqnarray}
where $C_{\mp}=(1\mp\alpha)(1-\gamma^{2})$, and the boundary lines $h_{d1}, h_{d2}$ are also called the disorder lines. As seen in Fig.~\ref{fig:phase.diagram}, the region inside the two dot lines is the incommensurate phase, and the outside is the commensurate phase. 

We study the non-equilibrium dynamical evolution induced by a quantum quench, in which the system is prepared in the ground state $|\psi_{0}\rangle=\bigotimes_{k>0}|\psi_{0k}\rangle, |\psi_{0k}\rangle=|\text{GS}_{k}\rangle$ of an initial Hamiltonian $H_{0}=H(h_{0},\gamma_{0})$. At time $t=0$, the system parameters are suddenly changed to $(h_{1},\gamma_{1})$ corresponding to the Hamiltonian $\tilde{H}=H(h_{1},\gamma_{1})$. The time-evolved state is given by
\begin{equation}\label{eq:time-evolved.state}
  \begin{split}
    |\psi_{k}(t)\rangle & = e^{-i\tilde{H}_{k}t}|\psi_{0k}\rangle \\
      & = \frac{e^{i(\tilde{\Lambda}_{k1}+\tilde{\Lambda}_{k2})t}}{\mathcal{N}}\prod_{\mu,\nu=1}^{2}[1     \\
      & + G_{k\mu,-k\nu}e^{-it(\tilde{\Lambda}_{k\mu}+
      \tilde{\Lambda}_{-k\nu})}\tilde{\eta}_{k\mu}^{\dag}\tilde{\eta}_{-k\nu}^{\dag}]|\tilde{\psi}_{0k}\rangle,
  \end{split}
\end{equation}
where $\mathcal{N}$ is the normalization coefficient, and $G$ is an antisymmetrical matrix depending on the parameters of the pre- and post-quench Hamiltonian (see Appendix~A).

\begin{figure}
  \centering
  \includegraphics[width=1.0\linewidth]{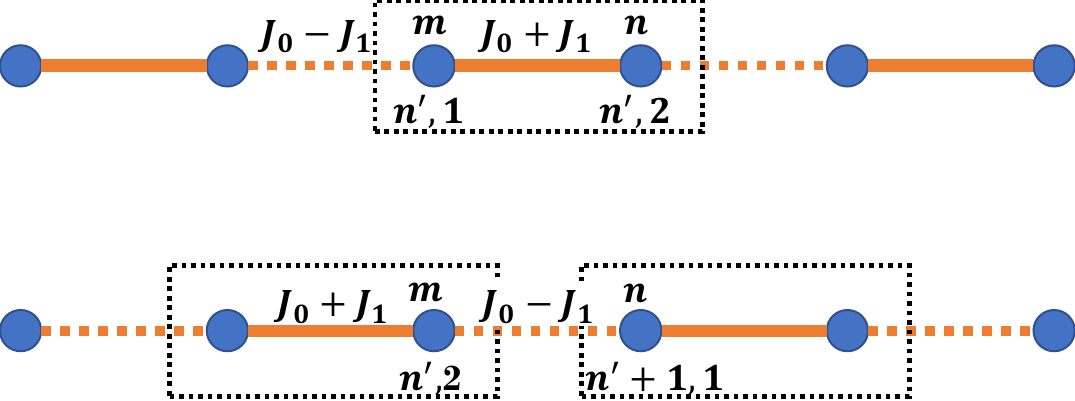}\\
  \caption{Sketches of two cases for the $C_{mn}(t)$. The indices $m(n)$ represent the lattice sites of the primary alternating chain, and the indices $(n',i), i=1,2$ represent the lattice sites after the system transformed to the double lattice.   }\label{fig:lattice.C.mn}
\end{figure}

To observe the dynamical relaxation behavior after a quench, we investigate the fermionic two-point correlation function $C_{mn}(t)=\langle\psi(t)|c_{m}^{\dag}c_{n}|\psi(t)\rangle$ following the Refs.~\onlinecite{PRB.2022.105.054301, PhysRevB.2023.108.014303}, where $|m-n|=1$ are considered. As illustrated in Fig.~\ref{fig:lattice.C.mn}, since the alternating XY model can be transformed to the spinless fermion model in the double lattice, $C_{mn}(t)$ has two different cases after the system mapped to the momentum space. When $C_{mn}(t)$ describes the correlation of two sites within the cell, it takes the form
\begin{equation}
  C_{mn}(t) = \langle\psi(t)|\frac{1}{N'}\sum_{k}c_{k1}^{\dag}c_{k2}|\psi(t)\rangle.
\end{equation}
While $C_{mn}(t)$ describes the correlation of two sites in the nearest-neighbor two cells, it takes the form
\begin{equation}
  C_{mn}(t) = \langle\psi(t)|\frac{1}{N'}\sum_{k}c_{k2}^{\dag}c_{k1}e^{ik}|\psi(t)\rangle.
\end{equation}
For both cases, we can obtain the expression of the deviation $\delta C_{mn}(t)=C_{mn}(t)-C_{mn}(\infty)$ through the analytical derivation and numerical verification (see Appendix~B). Specifically, for case 1, we have
\begin{equation}\label{eq:delta_Cmn.1}
  \begin{split}
    \delta C_{mn}(t) & = C_{mn}(t)-C_{mn}(\infty) \\
      & = \int_{0}^{\pi}\frac{dk}{2\pi}[f_{1}(k)\cos{[(\tilde{\Lambda}_{k1}-\tilde{\Lambda}_{k2})t]} \\
      & \quad\quad\quad
        + f_{2}(k)\cos{2\tilde{\Lambda}_{k1}t} + f_{3}(k)\cos{2\tilde{\Lambda}_{k2}t} \\
      & \quad\quad\quad
        + f_{4}(k)\cos{[(\tilde{\Lambda}_{k1}+\tilde{\Lambda}_{k2})t]}],
  \end{split}
\end{equation}
and for case 2, we have
\begin{equation}\label{eq:delta_Cmn.2}
  \begin{split}
    \delta C_{mn}(t) & = C_{mn}(t)-C_{mn}(\infty) \\
      & = \int_{0}^{\pi}\frac{dk}{2\pi}[f_{1}(k)\cos{[(\tilde{\Lambda}_{k1}-\tilde{\Lambda}_{k2})t]} \\
      & \quad\quad\quad
        + f_{2}(k)\cos{2\tilde{\Lambda}_{k1}t} + f_{3}(k)\cos{2\tilde{\Lambda}_{k2}t} \\
      & \quad\quad\quad
        + f_{4}(k)\cos{[(\tilde{\Lambda}_{k1}+\tilde{\Lambda}_{k2})t]}]\cos{k},
  \end{split}
\end{equation}
where the functions $f_{i}(k),i=1,\cdots,4$ are determined by the parameters of the pre- and post-quench Hamiltonian.
Considering that the behavior of $f_{\mu}(k)\cos{k}$ and $f_{\mu}(k)$ at the extreme points of $\Lambda_{k1}$ [see Eq.~(\ref{eq:extreme_point})] have no fundamental difference, Eq.~(\ref{eq:delta_Cmn.1}) and (\ref{eq:delta_Cmn.2}) are equivalent for discussing the relaxation behavior of $\delta C_{mn}(t)$.

\section{Dynamical scaling behavior of $\delta C_{mn}(t)$}

\begin{figure}
  \centering
  \includegraphics[width=1.0\linewidth]{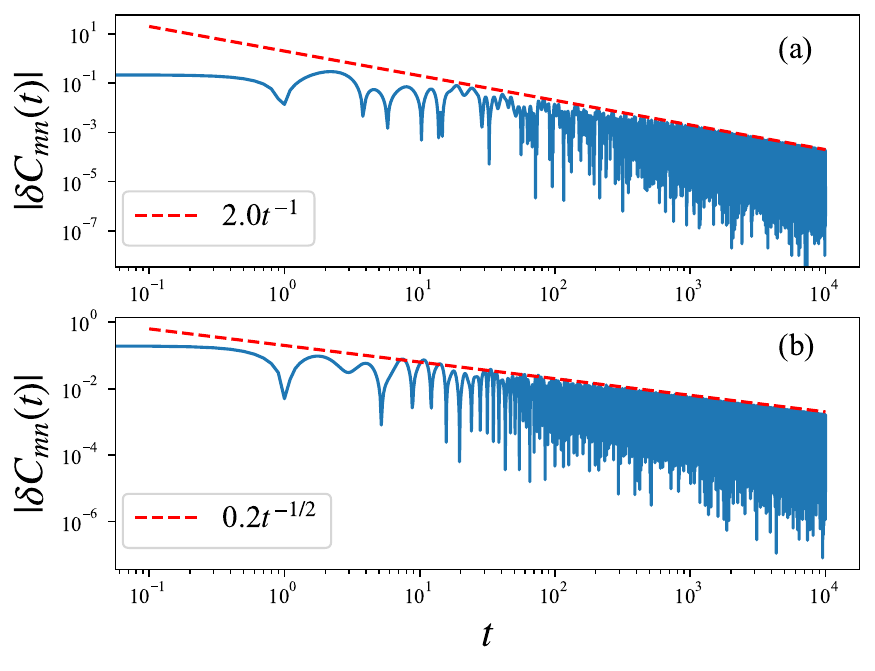}\\
  \caption{$|\delta C_{mn}(t)|$ as a function of $t$ for the quench (a) to the commensurate phase, and (b) to the incommensurate phase. }\label{fig:delta.Cmn}
\end{figure}

In this section, we present numerical results of the dynamical scaling behavior of $\delta C_{mn}(t)$ after the quench in the alternating XY chain. As seen in Fig.~\ref{fig:delta.Cmn}~(a) and (b), we display the typical results of $|\delta C_{mn}(t)|$ versing with $t$ for the quenches to the commensurate and incommensurate phases, respectively. The results reveal that the decay of $\delta C_{mn}(t)$ follows the scaling behavior $\sim t^{-1}$ for the quench to the commensurate phase, and $\sim t^{-1/2}$ for the quench to the incommensurate phase.

The relaxation behavior can be explained by the method of stationary point approximation. Unlike the case in the homogeneous XY chain, the deviation $\delta C_{mn}(t)$ can be divided into four components which can be considered independently. By using the Euler's formula, we can express $\delta C_{mn}(t)$ by
\begin{equation}
  \delta C_{mn}(t) = \text{Re}[\sum_{\mu=1}^{4}I_{\mu}(t)].
\end{equation}
Here, the function $I_{\mu}(t)$ has the integral form of the high-speed oscillation
\begin{equation}
  I_{\mu}(t) = \int_{0}^{\pi}\frac{dk}{2\pi}f_{\mu}(k)e^{ig_{\mu}(k)t},
\end{equation}
where the functions $g_{1}(k)=\tilde{\Lambda}_{k1}-\tilde{\Lambda}_{k1}$, $g_{2}(k)=2\tilde{\Lambda}_{k1}$, $g_{3}(k)=2\tilde{\Lambda}_{k2}$, and $g_{4}(k)=\tilde{\Lambda}_{k1}+\tilde{\Lambda}_{k2}$ are dependent on the quasiparticle excitation spectra. The exponential terms in $I_{\mu}(t)$ oscillates rapidly so that the values of $I_{\mu}(t)$ are determined by the integrals around the stationary points, which correspond to the extreme points of the functions $g_{\mu}(t)$.

\begin{figure}
  \centering
  \includegraphics[width=1.0\linewidth]{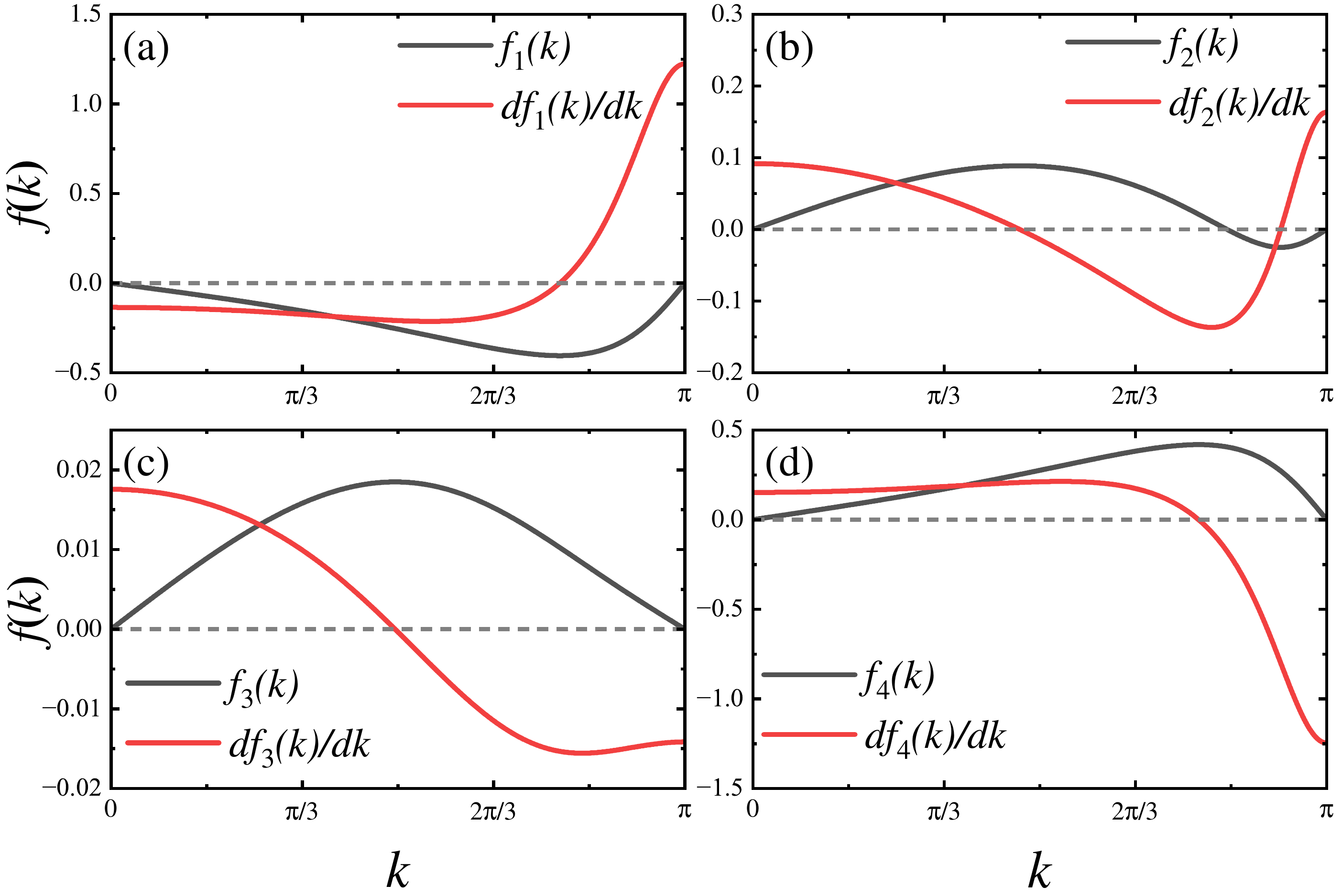}\\
  \caption{Numerical results of the functions $f_{1}(k)$, $f_{2}(k)$, $f_{3}(k)$, and $f_{4}(k)$ with their first-order derivatives for quenches to the commensurate phase.}\label{fig:f(k)}
\end{figure}

Take the integral of $I_{1}(t)$ as an example. For the quench to the commensurate phase, the function $g_{1}(t)=\tilde{\Lambda}_{k1}-\tilde{\Lambda}_{k2}$ has two extreme points, i.e. $k=0,\pi$. The value of $I_{1}(t)$ is then attributed by the integrals around the two stationary points $k=0,\pi$. However, according to the numerical results shown in Fig.~\ref{fig:f(k)}, the function $f_{1}(k)$ vanishes at both stationary points, but the first-order derivations $f_{1}'(0)$ and $f_{1}'(\pi)$ do not vanish. We thus expand $f_{1}(k)$ around two stationary points, which yields
\begin{eqnarray}
  f(k) &=& f_{1}'(0)k, \\
  f(k) &=& f_{1}'(\pi)(k-\pi).
\end{eqnarray}
Hence, we obtain the asymptotic behavior of $I_{1}(t)$ as
\begin{equation}
  \begin{split}
    I_{1}(t) & \approx \frac{1}{2}f_{1}'(0)e^{ig_{1}''(0)t}\int_{0}^{\infty} kdk e^{itg_{1}''(0)k^{2}} \\
      & + \frac{1}{2}f_{1}'(\pi)e^{ig_{1}''(\pi)t}\int_{0}^{\infty}(k-\pi)dk e^{itg_{1}''(\pi)(k-\pi)^{2}}.
  \end{split}
\end{equation}
By performing the Gaussian integral, we get the scaling relation
\begin{equation}
  I_{1}(t)\sim t^{-1}.
\end{equation}
Similarly, by considering that both $g_{2}(k), g_{3}(k)$ and $g_{4}(k)$ have the stationary points $k=0,\pi$, we can obtain the other three scaling relations
\begin{equation}
  I_{2}(t), I_{3}(t), I_{4}(t) \sim t^{-1}.
\end{equation}
Finally, the sum of four component follows the power-law decay of $\delta C_{mn}(t)\sim t^{-1}$, which agrees with the numerical results in Fig.~\ref{fig:delta.Cmn}.

For the quench to the incommensurate phase, we need to notice that according to Eq.~(\ref{eq:extreme_point}), the lower excitation spectrum $\Lambda_{k1}$ has three extreme points, i.e. $k=0,\pi$, and $k_{m}$ $(\tilde{\Lambda}_{k_{m},1}=\min{\tilde{\Lambda}_{k1}})$. It also should be noticed that $k_{m}$ is not the extreme point of the excitation spectrum $\tilde{\Lambda}_{k2}$. Therefore, among the functions $g_{\mu}(t)$, the function $g_{2}(k)=2\tilde{\Lambda}_{k1}$ has three stationary points ($k=0,\pi$, and $k_{m}$), but the other three functions $g_{1}(k), g_{3}(k)$ and $g_{4}(k)$ only have two stationary points ($k=0,\pi$). We thus can immediately obtain the scaling relations
\begin{equation}
  I_{1}(t), I_{3}(t), I_{4}(t)\sim t^{-1},
\end{equation}
for the quench to the incommensurate phase.

For the integral $I_{2}(t)$, it is known that the integrals around the two stationary points $k=0,\pi$ attribute the scaling relation $\sim t^{-1}$. However, for the third stationary point $k_{m}$, we have the integral
\begin{equation}
  \sim \frac{1}{2\pi}f_{2}(k_{m})e^{itg_{2}''(k_{m})}\int_{0}^{\pi} dk e^{itg_{2}''(k_{m})(k-k_{m})^{2}}.
\end{equation}
By performing the Gaussian integral, we get the contributions of the stationary point $k_{m}$ by $\sim t^{-1/2}$. For the long-time scale, the asymptotic behavior of $\delta C_{mn}(t)$ will be determined by the slowest decay. Therefore, for the quench to the incommensurate phase, the relaxation behavior of $\delta C_{mn}(t)$ is given by $\sim t^{-1/2}$, which agrees to the numerical results in Fig.~\ref{fig:delta.Cmn}~(b).

\subsection{Quench to the disorder lines}

\begin{figure}
  \centering
  \includegraphics[width=1.0\linewidth]{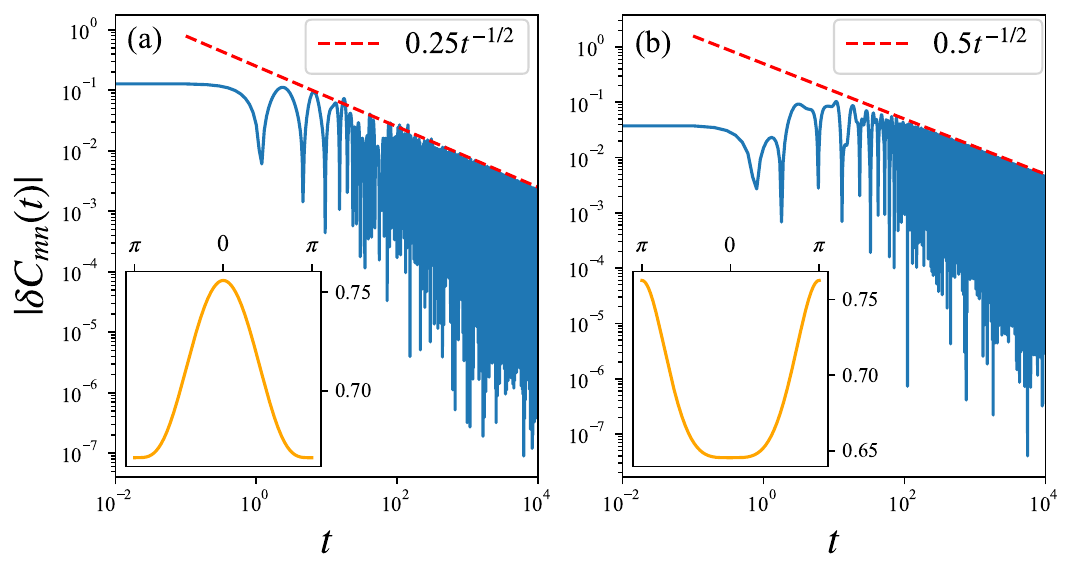}\\
  \caption{$|\delta C_{mn}(t)|$ as a function of $t$ for the quench to the disorder lines. (a) The quench is to the disorder line $h_{d1}$ [see the blue dot line in Fig.~\ref{fig:phase.diagram}]. The inset graph is the lower quasiparticle excitation spectrum $\tilde{\Lambda}_{k1}$ of the system on the disorder line $h_{d1}$. (b) The quench is to the disorder line $h_{d2}$ [see the red dot line in Fig.~\ref{fig:phase.diagram}]. The inset graph is the lower quasiparticle excitation spectrum $\tilde{\Lambda}_{k1}$ of the system on the disorder line $h_{d2}$.}\label{fig:quench.to.disorder.lines}
\end{figure}

Now, we consider a special quench protocol: quench to the disorder lines. In Fig.~\ref{fig:quench.to.disorder.lines}, we display two typical numerical results of $|\delta C_{mn}(t)|$ versing with $t$ for the quenches from the PM phase to the disorder line $h_{d1}$ [see the blue dot line in Fig.~\ref{fig:phase.diagram} and Eq.~(\ref{eq:disorder.line.1})], and from the commensurate part of FM$_x$ phase to the disorder line $h_{d2}$ [see the red dot line in Fig.~\ref{fig:phase.diagram} and Eq.~(\ref{eq:disorder.line.2})], respectively. It is observed that the scaling behavior is given by $\delta C_{mn} \sim t^{-1/2}$ for the quench to the disorder lines, which is the same as that in the quench to the incommensurate phase.

To explain the relaxation behavior for the quench protocol to the disorder lines, we need to emphasize that when the system parameters are on the disorder line, the solution that satisfies Eq.~(\ref{eq:extreme_point}) is the center of boundary of the Brillouin zone, i.e. $k_{m}=0$ or $\pi$. As seen in the inset graphs of Figs.~\ref{fig:quench.to.disorder.lines}~(a) and (b), we display the excitation spectrum $\tilde{\Lambda}_{k1}$ of the Hamiltonian with parameters on the disorder lines $h_{1}$ and $h_{2}$, respectively. It is observed that $k_{m}=\pi$ for the parameters on the disorder line $h_{1}$, and $k_{m}=0$ for the parameters on the disorder line $h_{2}$. Moreover, the first-order, second-order, and third-order derivations of $\Lambda_{k1}$ vanish at the wave vector $k_{m}$. Therefore, among the functions $g_{\mu}(t)$, the function $g_{2}(k)=2\tilde{\Lambda}_{k1}$ has one first-order stationary point and one third-order stationary point, but the three functions $g_{1}(k), g_{3}(k)$ and $g_{4}(k)$ only have two first-order stationary points. The asymptotic behaviors of $I_{1}(k), I_{3}(k)$ and $I_{4}(k)$ can be immediately obtained by
\begin{equation}
  I_{1}(t), I_{3}(t), I_{4}(t) \sim t^{-1}.
\end{equation}
However, for the integral $I_{2}(t)$, the contribution from the integral around the third-order stationary point takes the form (take the parameters located at the disorder line $h_{1}$ as an example, i.e. $k_{m}=\pi$)
\begin{equation}
  \sim \frac{1}{2}f_{2}'(\pi)\int_{0}^{\pi}(k-\pi)dk\exp{[i\frac{d^{4}g_{2}(k)}{dk^{4}}|_{k=\pi}(k-\pi)^{4}t]}.
\end{equation}
By performing the Gaussian integral, we get the contributions of the third-order stationary point by $\sim t^{-1/2}$. Therefore, for the quench to the disorder line, the relaxation behavior of $\delta C_{mn}(t)$ is given by $\sim t^{-1/2}$, which agrees to the numerical results in Fig.~\ref{fig:quench.to.disorder.lines}.

\section{Singularity behavior of $C_{mn}(\infty)$}

\begin{figure}
  \centering
  \includegraphics[width=1.0\linewidth]{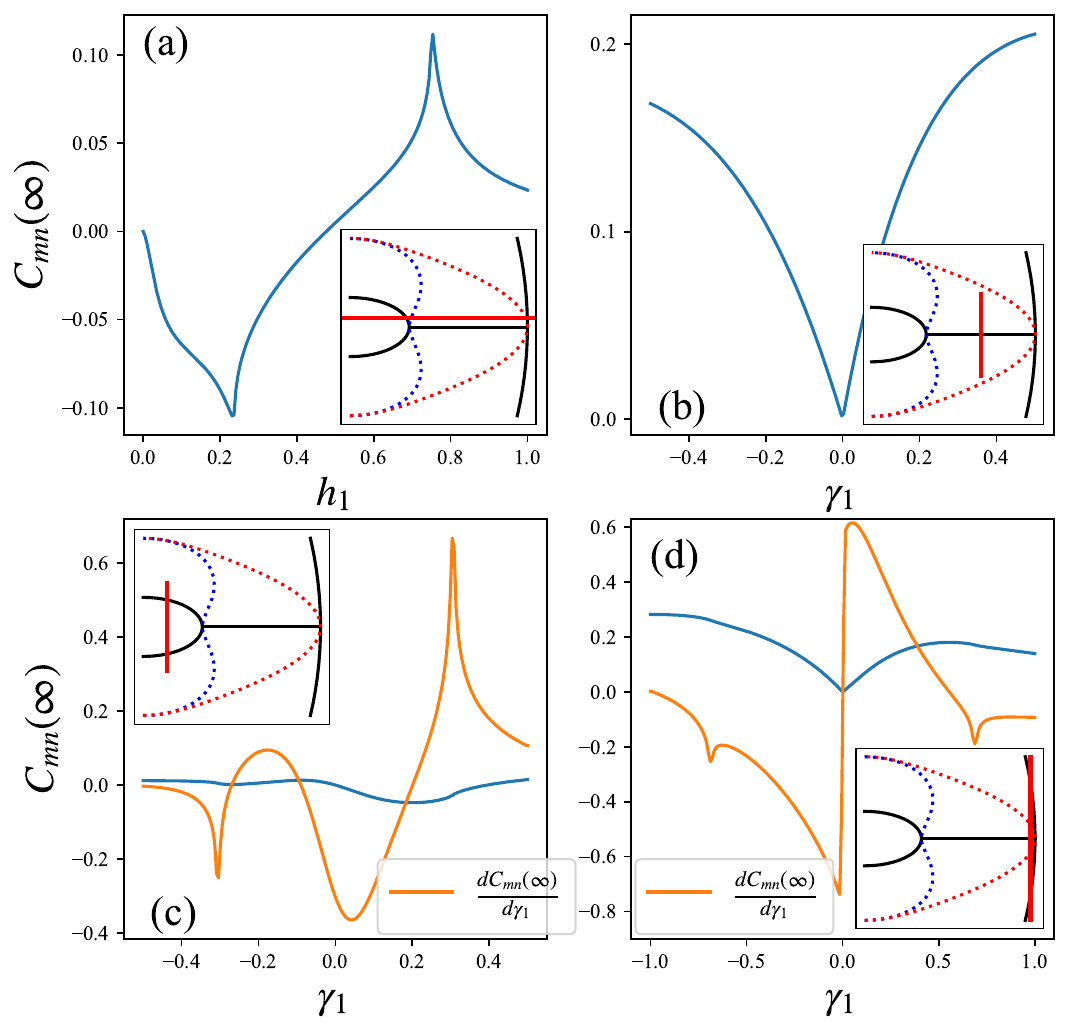}\\
  \caption{(a) The steady value $C_{mn}(\infty)$ as a function of the post-quench parameter $h_{1}$ for the quench to the line $\gamma_{1}=0.1$, corresponding to the quench crossing two Ising transitions successively. (b)-(d): $C_{mn}(\infty)$ as a function of the post-quench anisotropic parameter $\gamma_{1}$ for the quench (b) to the line $h_{1}=0.5$ (across the anisotropic transition), (c) to the line $h_{1}=0.1$ (across two Ising transitions), and (d) to the line $h_{1}=0.73$ (across Ising-anisotropic-Ising transitions). The red lines in the inset graphs are the target lines after quenches. The orange lines in (c) and (d) denote the first-order derivatives $\frac{dC_{mn}(\infty)}{d\gamma_{1}}$ of $C_{mn}(\infty)$ over the parameter $\gamma_{1}$. }\label{fig:Cmn.infty}
\end{figure}

In order to further understand the long-time relaxation behavior of the fermionic two-point correlation $C_{mn}(t)=C_{mn}(\infty)+\delta C_{mn}(t)$, it is not enough to only focus on the scaling behavior of the deviation $\delta C_{mn}(t)$. For this purpose, we analyze details of the steady value $C_{mn}(\infty)$ and the deviation $\delta C_{mn}(t)$. We obtain the expression of $C_{mn}(\infty)=\frac{1}{N'}\sum_{k>0}C_{mn}^{k}(\infty)$ in Appendix~B with
\begin{equation}
  \begin{split}
    C_{mn}^{k}(\infty) & =
          \tilde{Z}^{\dag}_{11}\tilde{Z}_{21}\langle\tilde{\eta}_{k1}^{\dag}\tilde{\eta}_{k1}\rangle_{t} 
         +\tilde{Z}_{31}\tilde{Z}_{14}^{\dag}\langle\tilde{\eta}_{k1}\tilde{\eta}_{k1}^{\dag}\rangle_{t} \\
      & 
         +\tilde{Z}^{\dag}_{21}\tilde{Z}_{22}\langle\tilde{\eta}_{k2}^{\dag}\tilde{\eta}_{k2}\rangle_{t} 
        +\tilde{Z}_{32}\tilde{Z}_{24}^{\dag}\langle\tilde{\eta}_{k2}\tilde{\eta}_{k2}^{\dag}\rangle_{t} \\
      & 
         +\tilde{Z}^{\dag}_{31}\tilde{Z}_{23}\langle\tilde{\eta}_{-k1}\tilde{\eta}_{-k1}^{\dag}\rangle_{t} 
        +\tilde{Z}_{33}\tilde{Z}_{34}^{\dag}\langle\tilde{\eta}_{-k1}^{\dag}\tilde{\eta}_{-k1}\rangle_{t} \\
      & 
        +\tilde{Z}^{\dag}_{41}\tilde{Z}_{24}\langle\tilde{\eta}_{-k2}\tilde{\eta}_{-k2}^{\dag}\rangle_{t} 
        +\tilde{Z}_{34}\tilde{Z}_{44}^{\dag}\langle\tilde{\eta}_{-k2}^{\dag}\tilde{\eta}_{-k2}\rangle_{t},
  \end{split}
\end{equation}
where the matrix $\tilde{Z}$ consists of the eigenvectors of the post-quench Hamiltonian, and $\langle.\rangle_{t}=\langle\psi(t)|.|\psi(t)\rangle$ denotes the average over the time-evolved states after a quench, instead of the ground states of the pre- and post-quench Hamiltonian. Therefore, the steady value after quench is different from the ground state expectation value \cite{JstatPhys.2009.135.599, Suzuki2013, PhysRevB.2010.82.144302, PhysRevA.1970.2.1075, PhysRevA.1971.3.786}. The steady value is found to only depend on the average fermion occupation numbers (i.e. the average single excited particle) [$\langle\tilde{\eta}_{k1}^{\dag}\tilde{\eta}_{k1}\rangle_{t}$, $\langle\tilde{\eta}_{k2}^{\dag}\tilde{\eta}_{k2}\rangle_{t}$, $\langle\tilde{\eta}_{-k1}^{\dag}\tilde{\eta}_{-k1}\rangle_{t}$, and $\langle\tilde{\eta}_{-k2}^{\dag}\tilde{\eta}_{-k2}\rangle_{t}$] which are exactly independent of the time.

\begin{figure}
  \centering
  \includegraphics[width=1.0\linewidth]{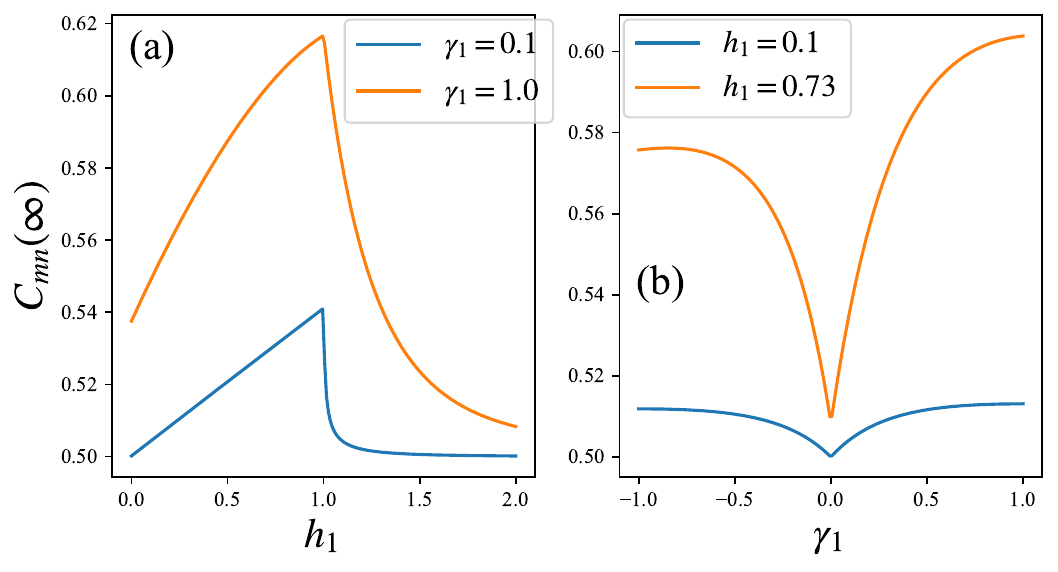}\\
  \caption{In the uniform XY chain, (a) The steady value $C_{mn}(\infty)$ as a function of the post-quench parameter $h_{1}$ for quenches to the lines $\gamma_{1}=0.1$ and $\gamma_{1}=1.0$, corresponding to quenches crossing the Ising transition $h_{c}=1$. (b) $C_{mn}(\infty)$ as a function of the post-quench anisotropic parameter $\gamma_{1}$ for quenches to the lines $h_{1}=0.1$ and $h_{1}=0.73$, corresponding to quenches crossing the anisotropic transition $\gamma_{c}=0$. }\label{fig:uniform.Cmn.infty}
\end{figure}

To examine more information on $C_{mn}(\infty)$, we study the steady value $C_{mn}(\infty)$ varying with the post-quench parameters. In Fig.~\ref{fig:Cmn.infty}, we display the steady values $C_{mn}(\infty)$ as functions of the post-quench parameters $h_{1}$ and $\gamma_{1}$, respectively. $C_{mn}(\infty)$ are found to have the finite value and exhibit the cusp-like singularities for the quench the two critical points of the Ising transitions $(h_{1}=h_{c})$ and the critical point of the anisotropic transition $(\gamma_{c}=0)$ [see Fig.~\ref{fig:Cmn.infty}~(a) and (b)]. Meanwhile, for the quench crossing the Ising transition with fixed external field $h_{1}$, $C_{mn}(\infty)$ shows the inflection points for the quench to the critical points $(\gamma_{1}=\gamma_{c})$, at which the first-order derivatives $\frac{dC_{mn}(\infty)}{d\gamma_{1}}$ exhibit the cusp-like singularities [see Fig.~\ref{fig:Cmn.infty}~(c) and (d)]. In addition, we also investigate the singularity behavior of $C_{mn}(\infty)$ of the uniform XY chain, as shown in Fig.~\ref{fig:uniform.Cmn.infty}, which was not previously explored in the literature \cite{PhysRevA.2017.95.023621, PhysRevB.2019.100.115150, PRB.2022.105.054301, PhysRevB.2023.107.075138, PhysRevB.2023.108.014303}. Our results show that $C_{mn}(\infty)$ shows the cusp-like singularities for the quench to the critical points of the Ising transition $h_{1}=h_{c}=1$ and the anisotropic transition $\gamma_{1}=\gamma_{c}=0$ of the uniform XY chain. These findings provide insight into the relationship between the steady value $C_{mn}(\infty)$ and the QPTs of the alternating and uniform XY chain.

On the other hand, based on our derivation [see Eq.~(\ref{eq:deltaC.1}) in Appendix~B], we can also identify the origin of the deviation $\delta C_{mn}(t)$. In the alternating XY chain, the decay of $\delta C_{mn}(t)$ originates from the transients of two aspects of physical processes. One is corresponding to the double excited quasiparticle excitation, i.e. $\langle\tilde{\eta}_{k1}^{\dag}\tilde{\eta}_{-k1}^{\dag}\rangle_{t}$, $\langle\tilde{\eta}_{k1}^{\dag}\tilde{\eta}_{-k2}^{\dag}\rangle_{t}$, and $\langle\tilde{\eta}_{-k1}\tilde{\eta}_{k1}\rangle_{t}$ etc. In fact, this perspective can also be inferred from the method of solving the time-dependent Bogoliubov-de Gennes equation in the homogeneous XY chain \cite{PRB.2022.105.054301,PhysRevB.2023.108.014303}, where the BCS-type state at all times is the ansatz
\begin{equation}
  |\psi(t)\rangle = \prod_{k>0}[u(t)+v(t)c_{k}^{\dag}c_{-k}^{\dag}]|0\rangle.
\end{equation}
It is observed that the deviation $\delta C_{mn}(t)$ depends on the time-dependent part of $v(t)$. Therefore, these terms are related to the superconducting terms $(c_{k\mu}^{\dag}c_{-k\nu}^{\dag}+h.c.)$ in the spinless fermion model which is transformed from the XY chain via the Jordan-Wigner transformation. Another one is corresponding to the average quasiparticle transition terms between different excitation spectra, i.e. $\langle\tilde{\eta}_{k1}^{\dag}\tilde{\eta}_{k2}\rangle_{t}$, $\langle\tilde{\eta}_{k2}^{\dag}\tilde{\eta}_{k1}\rangle_{t}$, $\langle\tilde{\eta}_{-k1}\tilde{\eta}_{-k2}^{\dag}\rangle_{t}$, $\langle\tilde{\eta}_{-k2}\tilde{\eta}_{-k1}^{\dag}\rangle_{t}$. Since the alternating XY chain has two quasiparticle excitation spectra, the nonequilibrium induced by the quench causes excited quasiparticles to transition between two excitation spectra, and this effect vanishes for a long time. As a result, these transition terms can not be seen in the uniform XY model due to its multi-band nature.

\section{Conclusion}

In this paper, we study the relaxation dynamics of the fermion two-point correlation function in the XY chain with the alternating nearest-neighbor hopping interaction after a quench. We find that the inhomogeneity induced by the alternating hopping interaction can affect the scaling behavior of the deviation $\delta C_{mn}(t)$. Specifically, when quenching to the incommensurate phase and the disorder lines, $\delta C_{mn}(t)$ decays with time following the power law behavior of $\sim t^{-1/2}$. While the quench is to the commensurate phase, $\delta C_{mn}(t)$ decays with $t^{-1}$, which is different from the case of the uniform XY model. In the uniform XY model, $\delta C_{mn}(t)$ decays with time by $t^{-3/4}$, and $t^{-3/2}$ for the quench to the disorder line and the commensurate phase \cite{PRB.2022.105.054301}, respectively. The reason for the change of scaling behavior in our model can be explained using stationary point approximate methods. Moreover, based on our derivation of $\delta C_{mn}(t)$, we determine that the decay of $\delta C_{mn}(t)$ originates from the transients of the double-excited quasiparticle excitation and the transitions between different excitation spectra. The former is related to the superconducting effect in the quantum spin model, and the latter is related to the multi-band effect of the alternating XY chain. In addition, we examine the underlying physics of the correlation function converging to the steady value $C_{mn}(\infty)$. By obtaining the expression of $C_{mn}(\infty)$, we find that the steady value only involves the average fermion occupation numbers which are independent of the time. Since the time-evolved state is not the ground state of the pre- and post-quench Hamiltonian, the steady value after a quench is different from the ground state expectation value \cite{JstatPhys.2009.135.599, Suzuki2013, PhysRevB.2010.82.144302, PhysRevA.1970.2.1075, PhysRevA.1971.3.786}. It is found that the steady value exhibits the nonanalytic singularities for the quench to the critical points of the QPTs, which suggests $C_{mn}(\infty)$ to characterize the QPT. The generality of the singular behavior of the steady value is also confirmed in the uniform XY chain.

\begin{acknowledgments}
 The work is supported by the National Science Foundation of China (Grant Nos.~12204432, 11975126, 12204432), and Key Research Projects of Zhejiang Lab (Nos. 2021PB0AC01 and 2021PB0AC02).
\end{acknowledgments}

\appendix

\section{Quantum quench in the alternating XY chain}

By applying the Jordan-Wigner transformation, the Hamiltonian (\ref{eq:Hamil.XY.alternating}) of the alternating XY chain can be mapped into the spinless fermion model in the double lattice by
\begin{equation}
  \begin{split}
    H & = -\frac{1}{2}\sum_{n=1}^{N'}\{[(c_{n,1}^{\dag}c_{n,2}+c_{n,1}^{\dag}c_{n,2}^{\dag})+hc_{n,1}^{\dag}c_{n,1}] \\
      & +[\alpha (c_{n,2}^{\dag}c_{n+1,1}+c_{n,2}^{\dag}c_{n+1,1}^{\dag})+hc_{n,2}^{\dag}c_{n,2}]+h.c.\},
  \end{split}
\end{equation}
where $N'=N/2$. By performing the Fourier transformation in unit cell with $c_{n,\mu}=\frac{1}{\sqrt{N'}}\sum_{k}e^{ikn}c_{k\mu}$ and $c_{n,\mu}^{\dag}=\frac{1}{\sqrt{N'}}\sum_{k}e^{-ikn}c_{k\mu}^{\dag}$, we obtain the Hamiltonian in momentum space by
\begin{equation}
  H = \sum_{k>0}\Psi_{k}^{\dag}\mathbb{H}_{k}\Psi_{k},
\end{equation}
where the spinor operator is $\Psi_{k}^{\dag}=(\Gamma_{k}^{\dag}, \Gamma_{-k}^{T})$ with $\Gamma_{k}^{\dag}=(c_{k1}^{\dag}, c_{k2}^{\dag})$ and $\Gamma_{-k}^{T}=(c_{-k1}, c_{-k2})$, and the Bloch Hamiltonian $\mathbb{H}_{k}$ is given by
\begin{widetext}
  \begin{equation}
    \mathbb{H}_{k}=\frac{1}{2}\left(\begin{array}{cccc}
                -2h & 0 & -(1+\alpha e^{-ik}) & -\gamma(1-\alpha e^{-ik}) \\
                0 & 2h & \gamma(1-\alpha e^{-ik}) & (1+\alpha e^{-ik}) \\
                -(1+\alpha e^{ik}) & \gamma(1-\alpha e^{ik}) & -2h & 0 \\
                -\gamma(1-\alpha e^{ik}) & (1+\alpha e^{ik}) & 0 & 2h \\
                \end{array}\right).
  \end{equation}
\end{widetext}
The Bloch Hamiltonian $\mathbb{H}_{k}$ can be diagonalized by the standard diagonalization procedure of the Hermitian matrix, where we obtain $\mathbb{H}_{k} = Z\Lambda_{k}Z^{\dag}$ with $\Lambda_{k}=\text{diag}(\Lambda_{k1},\Lambda_{k2},-\Lambda_{-k1},-\Lambda_{-k2})$. By defining the canonical transformation
\begin{equation}\label{eq:canonical.transformation}
  (\Phi_{k}^{\dag}, \Phi_{-k}^{T}) = (\Gamma_{k}^{\dag}, \Gamma_{-k}^{T})Z,
\end{equation}
with $\Phi_{k}^{\dag}=(\eta_{k1}^{\dag}, \eta_{k2}^{\dag})$, and $\Phi_{-k}^{T}=(\eta_{-k1}, \eta_{-k2})$, we obtain the Hamiltonian in the diagonal form
\begin{equation}
  H = \sum_{k}[\Lambda_{k1}(\eta_{k1}^{\dag}\eta_{k1}-\frac{1}{2})+\Lambda_{k2}(\eta_{k2}^{\dag}\eta_{k2}-\frac{1}{2})].
\end{equation}

The canonical transformation (\ref{eq:canonical.transformation}) is actually corresponding the Bogoliubov transformation, which can also be expressed as
\begin{equation}
  \left(
    \begin{array}{l}
      \Phi_{k} \\
      \Phi_{-k} \\
      \Phi_{k}^{\dag T} \\
      \Phi_{-k}^{\dag T} \\
    \end{array}
  \right) = \left(
              \begin{array}{cc}
                U & V \\
                V^{*} & U^{*} \\
              \end{array}
            \right)\left(
    \begin{array}{l}
      \Gamma_{k} \\
      \Gamma_{-k} \\
      \Gamma_{k}^{\dag T} \\
      \Gamma_{-k}^{\dag T} \\
    \end{array}
  \right) = M\left(
    \begin{array}{l}
      \Gamma_{k} \\
      \Gamma_{-k} \\
      \Gamma_{k}^{\dag T} \\
      \Gamma_{-k}^{\dag T} \\
    \end{array}
  \right).
\end{equation}
The Bogoliubov coefficient matrices $U$ and $V$ can be obtained by
\begin{equation}
  U = \left(
        \begin{array}{cc}
          Z_{11}^{\dag} & 0 \\
          0 & Z_{22}^{T} \\
        \end{array}
      \right)
\end{equation}
and
\begin{equation}
  V = \left(
        \begin{array}{cc}
          0 & Z_{21}^{\dag} \\
          Z_{12}^{T} & 0 \\
        \end{array}
      \right)
\end{equation}
with $Z=\left(
          \begin{array}{cc}
            Z_{11} & Z_{12} \\
            Z_{21} & Z_{22} \\
          \end{array}
        \right)
$.

In a quantum quench, we prepare the ground states $|\psi_{0}\rangle = \bigotimes_{k>0}|\psi_{0k}\rangle$ of the Hamiltonian $H(h_{0})$ as the initial state, and suddenly change the system into the Hamiltonian $\tilde{H} = H(h_{1})$ at time $t = 0$. From the relation between the Bogoliubov quasi-particle operators
\begin{equation}
    \left(
    \begin{array}{l}
      \Phi_{k} \\
      \Phi_{-k} \\
      \Phi_{k}^{\dag T} \\
      \Phi_{-k}^{\dag T} \\
    \end{array}
  \right)  = M\tilde{M}^{-1}
    \left(
    \begin{array}{l}
      \tilde{\Phi}_{k} \\
      \tilde{\Phi}_{-k} \\
      \tilde{\Phi}_{k}^{\dag T} \\
      \tilde{\Phi}_{-k}^{\dag T} \\
    \end{array}
  \right)
\end{equation}
with
\begin{equation}
  M\tilde{M}^{-1} = \left(
           \begin{array}{cc}
             U\tilde{U}^{\dag}+V\tilde{V}^{\dag} & U\tilde{V}^{T}+V\tilde{U}^{T} \\
             U^{*}\tilde{V}^{\dag}+V^{*}\tilde{U}^{\dag} & U^{*}\tilde{U}^{T}+V^{*}\tilde{V}^{T} \\
           \end{array}
         \right).
\end{equation}
From the equations $\eta_{k}|\psi_{0k}\rangle=0$ and $\tilde{\eta}_{k}|\tilde{\psi}_{0k}\rangle=0$, the initial state hence can be given by \cite{PhysRevLett.2008.101.120603}
\begin{equation}\label{initial.states}
  |\psi_{0k}\rangle = \frac{1}{\mathcal{N}}\exp{[\frac{1}{2}
  \left(
    \begin{array}{cc}
      \tilde{\Phi}_{k}^{\dag} & \tilde{\Phi}_{-k}^{\dag} \\
    \end{array}
  \right)G
  \left(
    \begin{array}{c}
      \tilde{\Phi}_{k}^{\dag T} \\
      \tilde{\Phi}_{-k}^{\dag T}
    \end{array}
  \right)
  ]}|\tilde{\psi}_{0k}\rangle,
\end{equation}
with $G=-(U\tilde{U}^{\dag}+V\tilde{V}^{\dag})^{-1}(U\tilde{V}^{T}+V\tilde{U}^{T})$ \cite{PhysRevA.2011.84.052105}.

\begin{widetext}

\section{Two-point correlation function after a quench}

Since the alternating XY chain can be transformed into the spinless fermion model in the double-sublattice model (see Fig.~\ref{fig:lattice.C.mn}), we obtain the expression of $C_{mn}(t)$ in momentum space by ($|m-n|=1$)
\begin{equation}
  C_{mn}(t) = \langle\psi(t)|\frac{1}{N'}\sum_{k}c_{k1}^{\dag}c_{k2}|\psi(t)\rangle
\end{equation}
for Case 1, and
\begin{equation}
  C_{mn}(t) = \langle\psi(t)|\frac{1}{N'}\sum_{k}c_{k2}^{\dag}c_{k1}e^{ik}|\psi(t)\rangle
\end{equation}
for Case 2.

For Case 1, we have
\begin{equation}\label{eq:C_mn.case1}
    C_{mn}(t) = \frac{1}{N'}\sum_{k>0}[\langle\psi_{k}(t)|c_{k1}^{\dag}c_{k2}|\psi_{k}(t)\rangle + \langle\psi_{k}(t)|c_{-k1}^{\dag}c_{-k2}|\psi_{k}(t)\rangle].
\end{equation}
According to Eq.~(\ref{eq:canonical.transformation}), we have
\begin{eqnarray}
  c_{k1}^{\dag} &=& \tilde{Z}^{\dag}_{11}\tilde{\eta}_{k1}^{\dag}+\tilde{Z}^{\dag}_{21}\tilde{\eta}_{k2}^{\dag}+\tilde{Z}^{\dag}_{31}\tilde{\eta}_{-k1}
  +\tilde{Z}^{\dag}_{41}\tilde{\eta}_{-k2}, \\
  c_{k2} &=& \tilde{Z}_{21}\tilde{\eta}_{k1}+\tilde{Z}_{22}\tilde{\eta}_{k2}+\tilde{Z}_{23}\tilde{\eta}_{-k1}^{\dag}
  +\tilde{Z}_{24}\tilde{\eta}_{-k2}^{\dag}, \\
  c_{-k1}^{\dag} &=& \tilde{Z}_{31}\tilde{\eta}_{k1}+\tilde{Z}_{32}\tilde{\eta}_{k2}+\tilde{Z}_{33}\tilde{\eta}_{-k1}^{\dag}
  +\tilde{Z}_{34}\tilde{\eta}_{-k2}^{\dag}, \\
  c_{-k2} &=& \tilde{Z}^{\dag}_{14}\tilde{\eta}_{k1}^{\dag}+\tilde{Z}^{\dag}_{24}\tilde{\eta}_{k2}^{\dag}+\tilde{Z}^{\dag}_{34}\tilde{\eta}_{-k1}
  +\tilde{Z}^{\dag}_{44}\tilde{\eta}_{-k2}.
\end{eqnarray}
Then, the first term of (\ref{eq:C_mn.case1}) is given by
\begin{equation}
  \begin{split}
    \langle c_{k1}^{\dag}c_{k2}\rangle_{t}
      & =
      \tilde{Z}^{\dag}_{11}\tilde{Z}_{21}\langle\tilde{\eta}_{k1}^{\dag}\tilde{\eta}_{k1}\rangle_{t} + \tilde{Z}^{\dag}_{11}\tilde{Z}_{22}\langle\tilde{\eta}_{k1}^{\dag}\tilde{\eta}_{k2}\rangle_{t}  + \tilde{Z}^{\dag}_{11}\tilde{Z}_{23}\langle\tilde{\eta}_{k1}^{\dag}\tilde{\eta}_{-k1}^{\dag}\rangle_{t}  + \tilde{Z}^{\dag}_{11}\tilde{Z}_{24}\langle\tilde{\eta}_{k1}^{\dag}\tilde{\eta}_{-k2}^{\dag}\rangle_{t} \\
      & +
      \tilde{Z}^{\dag}_{21}\tilde{Z}_{21}\langle\tilde{\eta}_{k2}^{\dag}\tilde{\eta}_{k1}\rangle_{t} + \tilde{Z}^{\dag}_{21}\tilde{Z}_{22}\langle\tilde{\eta}_{k2}^{\dag}\tilde{\eta}_{k2}\rangle_{t}  + \tilde{Z}^{\dag}_{21}\tilde{Z}_{23}\langle\tilde{\eta}_{k2}^{\dag}\tilde{\eta}_{-k1}^{\dag}\rangle_{t}  + \tilde{Z}^{\dag}_{21}\tilde{Z}_{24}\langle\tilde{\eta}_{k2}^{\dag}\tilde{\eta}_{-k2}^{\dag}\rangle_{t} \\
      & +
      \tilde{Z}^{\dag}_{31}\tilde{Z}_{21}\langle\tilde{\eta}_{-k1}\tilde{\eta}_{k1}\rangle_{t} + \tilde{Z}^{\dag}_{31}\tilde{Z}_{22}\langle\tilde{\eta}_{-k1}\tilde{\eta}_{k2}\rangle_{t}  + \tilde{Z}^{\dag}_{31}\tilde{Z}_{23}\langle\tilde{\eta}_{-k1}\tilde{\eta}_{-k1}^{\dag}\rangle_{t}  + \tilde{Z}^{\dag}_{31}\tilde{Z}_{24}\langle\tilde{\eta}_{-k1}\tilde{\eta}_{-k2}^{\dag}\rangle_{t} \\
      & +
      \tilde{Z}^{\dag}_{41}\tilde{Z}_{21}\langle\tilde{\eta}_{-k2}\tilde{\eta}_{k1}\rangle_{t} + \tilde{Z}^{\dag}_{41}\tilde{Z}_{22}\langle\tilde{\eta}_{-k2}\tilde{\eta}_{k2}\rangle_{t}  + \tilde{Z}^{\dag}_{41}\tilde{Z}_{23}\langle\tilde{\eta}_{-k2}\tilde{\eta}_{-k1}^{\dag}\rangle_{t}  + \tilde{Z}^{\dag}_{41}\tilde{Z}_{24}\langle\tilde{\eta}_{-k2}\tilde{\eta}_{-k2}^{\dag}\rangle_{t},
  \end{split}
\end{equation}
and
\begin{equation}
  \begin{split}
    \langle c_{-k1}^{\dag}c_{-k2}\rangle_{t}
      & =
      \tilde{Z}_{31}\tilde{Z}_{14}^{\dag}\langle\tilde{\eta}_{k1}\tilde{\eta}_{k1}^{\dag}\rangle_{t} +
      \tilde{Z}_{31}\tilde{Z}_{24}^{\dag}\langle\tilde{\eta}_{k1}\tilde{\eta}_{k2}^{\dag}\rangle_{t} +
      \tilde{Z}_{31}\tilde{Z}_{34}^{\dag}\langle\tilde{\eta}_{k1}\tilde{\eta}_{-k1}\rangle_{t} +
      \tilde{Z}_{31}\tilde{Z}_{44}^{\dag}\langle\tilde{\eta}_{k1}\tilde{\eta}_{-k2}\rangle_{t} \\
      & +
      \tilde{Z}_{32}\tilde{Z}_{14}^{\dag}\langle\tilde{\eta}_{k2}\tilde{\eta}_{k1}^{\dag}\rangle_{t} +
      \tilde{Z}_{32}\tilde{Z}_{24}^{\dag}\langle\tilde{\eta}_{k2}\tilde{\eta}_{k2}^{\dag}\rangle_{t} +
      \tilde{Z}_{32}\tilde{Z}_{34}^{\dag}\langle\tilde{\eta}_{k2}\tilde{\eta}_{-k1}\rangle_{t} +
      \tilde{Z}_{32}\tilde{Z}_{44}^{\dag}\langle\tilde{\eta}_{k2}\tilde{\eta}_{-k2}\rangle_{t} \\
      & +
      \tilde{Z}_{33}\tilde{Z}_{14}^{\dag}\langle\tilde{\eta}_{-k1}^{\dag}\tilde{\eta}_{k1}^{\dag}\rangle_{t} +
      \tilde{Z}_{33}\tilde{Z}_{24}^{\dag}\langle\tilde{\eta}_{-k1}^{\dag}\tilde{\eta}_{k2}^{\dag}\rangle_{t} +
      \tilde{Z}_{33}\tilde{Z}_{34}^{\dag}\langle\tilde{\eta}_{-k1}^{\dag}\tilde{\eta}_{-k1}\rangle_{t} +
      \tilde{Z}_{33}\tilde{Z}_{44}^{\dag}\langle\tilde{\eta}_{-k1}^{\dag}\tilde{\eta}_{-k2}\rangle_{t} \\
      & +
      \tilde{Z}_{34}\tilde{Z}_{14}^{\dag}\langle\tilde{\eta}_{-k2}^{\dag}\tilde{\eta}_{k1}^{\dag}\rangle_{t} +
      \tilde{Z}_{34}\tilde{Z}_{24}^{\dag}\langle\tilde{\eta}_{-k2}^{\dag}\tilde{\eta}_{k2}^{\dag}\rangle_{t} +
      \tilde{Z}_{34}\tilde{Z}_{34}^{\dag}\langle\tilde{\eta}_{-k2}^{\dag}\tilde{\eta}_{-k1}\rangle_{t} +
      \tilde{Z}_{34}\tilde{Z}_{44}^{\dag}\langle\tilde{\eta}_{-k2}^{\dag}\tilde{\eta}_{-k2}\rangle_{t}
    \end{split}
\end{equation}
where $\langle.\rangle_{t}=\langle\psi_{k}(t)|.|\psi_{k}(t)\rangle$. According to Eq.~(\ref{eq:time-evolved.state}), we have
\begin{eqnarray}
  \langle\tilde{\eta}_{k1}^{\dag}\tilde{\eta}_{k1}\rangle_{t} &=& |G_{k1,-k1}|^{2}+|G_{k1,-k2}|^{2}+|G_{k1,-k1}G_{k2,-k2}-G_{k1,-k2}G_{k2,-1}|^{2}, \\
  \langle\tilde{\eta}_{k1}^{\dag}\tilde{\eta}_{k2}\rangle_{t} = \langle\tilde{\eta}_{k2}^{\dag}\tilde{\eta}_{k1}\rangle_{t}^{*} &=& (G_{k1,-k1}^{*}G_{k2,-k1}+G_{k1,-k2}^{*}G_{k2,-k2})e^{it(\tilde{\Lambda}_{k1}-\tilde{\Lambda}_{k2})}, \\
  \langle\tilde{\eta}_{k1}^{\dag}\tilde{\eta}_{-k1}^{\dag}\rangle_{t} = \langle\tilde{\eta}_{-k1}\tilde{\eta}_{k1}\rangle_{t}^{*} &=& [G_{k1,-k1}^{*}+(G_{k1,-k1}G_{k2,-k2}-G_{k1,-k2}G_{k2,-k1})^{*}G_{k2,-k2}]e^{it(\tilde{\Lambda}_{k1}+\tilde{\Lambda}_{-k1})}, \\
  \langle\tilde{\eta}_{k1}^{\dag}\tilde{\eta}_{-k2}^{\dag}\rangle_{t} = \langle\tilde{\eta}_{-k2}\tilde{\eta}_{k1}\rangle_{t}^{*} &=& [G_{k1,-k2}^{*}-(G_{k1,-k1}G_{k2,-k2}-G_{k1,-k2}G_{k2,-k1})^{*}G_{k2,-k1}]e^{it(\tilde{\Lambda}_{k1}+\tilde{\Lambda}_{-k2})}, \\
  \langle\tilde{\eta}_{k2}^{\dag}\tilde{\eta}_{k2}\rangle_{t} &=& |G_{k2,-k1}|^{2}+|G_{k2,-k2}|^{2}+|G_{k1,-k1}G_{k2,-k2}-G_{k1,-k2}G_{k2,-k1}|^{2}, \\
  \langle\tilde{\eta}_{k2}^{\dag}\tilde{\eta}_{-k1}^{\dag}\rangle_{t} = \langle\tilde{\eta}_{-k1}\tilde{\eta}_{k2}\rangle_{t}^{*} &=& [G_{k2,-k1}^{*}-(G_{k1,-k1}G_{k2,-k2}-G_{k1,-k2}G_{k2,-k1})^{*}G_{k1,-k2}]e^{it(\tilde{\Lambda}_{k2}+\tilde{\Lambda}_{-k1})}, \\
  \langle\tilde{\eta}_{k2}^{\dag}\tilde{\eta}_{-k2}^{\dag}\rangle_{t} = \langle\tilde{\eta}_{-k2}\tilde{\eta}_{k2}\rangle_{t}^{*} &=& [G_{k2,-k2}^{*}+(G_{k1,-k1}G_{k2,-k2}-G_{k1,-k2}G_{k2,-k1})^{*}G_{k1,-k1}]e^{it(\tilde{\Lambda}_{k2}+\tilde{\Lambda}_{-k2})}, \\
  \langle\tilde{\eta}_{-k1}\tilde{\eta}_{-k1}^{\dag}\rangle_{t} &=& 1+|G_{k1,-k2}|^{2}+|G_{k2,-k2}|^{2}, \\
  \langle\tilde{\eta}_{-k1}\tilde{\eta}_{-k2}^{\dag}\rangle_{t} = \langle\tilde{\eta}_{-k2}\tilde{\eta}_{-k1}^{\dag}\rangle_{t}^{*} &=& -(G_{k1,-k2}^{*}G_{k1,-k1}+G_{k2,-k2}^{*}G_{k2,-k1})e^{it(\tilde{\Lambda}_{-k2}-\tilde{\Lambda}_{-k1})}, \\
  \langle\tilde{\eta}_{-k2}\tilde{\eta}_{-k2}^{\dag}\rangle_{t} &=& 1+|G_{k1,-k1}|^{2}+|G_{k2,-k1}|^{2}.
\end{eqnarray}
It is clear that the difference $\delta C_{mn}(t)$ of the fermion two-point correlation function $C_{mn}(t)$ from its steady state value $C_{mn}(\infty)$ is contributed by the components $\langle\tilde{\eta}_{k\mu}^{\dag}\tilde{\eta}_{-k\nu}^{\dag}\rangle_{t}$ and $\langle\tilde{\eta}_{k\mu}\tilde{\eta}_{-k\nu}\rangle_{t}, \mu, \nu = 1, 2$. We obtain
\begin{equation}\label{eq:deltaC.1}
  \begin{split}
    \delta C_{mn}(t) & = C_{mn}(t)-C_{mn}(\infty) \\
      & = \frac{1}{N'}\sum_{k>0}[
          (\tilde{Z}^{\dag}_{11}\tilde{Z}_{22}-
          \tilde{Z}_{32}\tilde{Z}_{14}^{\dag}) \langle\tilde{\eta}_{k1}^{\dag}\tilde{\eta}_{k2}\rangle_{t} + (\tilde{Z}^{\dag}_{21}\tilde{Z}_{21}-
          \tilde{Z}_{31}\tilde{Z}_{24}^{\dag}) \langle\tilde{\eta}_{k2}^{\dag}\tilde{\eta}_{k1}\rangle_{t} \\
      & \quad\quad\quad
        + (\tilde{Z}^{\dag}_{11}\tilde{Z}_{23}-\tilde{Z}_{33}\tilde{Z}_{14}^{\dag}) \langle\tilde{\eta}_{k1}^{\dag}\tilde{\eta}_{-k1}^{\dag}\rangle_{t}
        + (\tilde{Z}^{\dag}_{31}\tilde{Z}_{21}-\tilde{Z}_{31}\tilde{Z}_{34}^{\dag}) \langle\tilde{\eta}_{-k1}\tilde{\eta}_{k1}\rangle_{t} \\
      & \quad\quad\quad
        + (\tilde{Z}^{\dag}_{11}\tilde{Z}_{24}-\tilde{Z}_{34}\tilde{Z}_{14}^{\dag}) \langle\tilde{\eta}_{k1}^{\dag}\tilde{\eta}_{-k2}^{\dag}\rangle_{t}
        + (\tilde{Z}^{\dag}_{41}\tilde{Z}_{21}-\tilde{Z}_{31}\tilde{Z}_{44}^{\dag}) \langle\tilde{\eta}_{-k2}\tilde{\eta}_{k1}\rangle_{t} \\
      & \quad\quad\quad
        + (\tilde{Z}^{\dag}_{21}\tilde{Z}_{23}-\tilde{Z}_{33}\tilde{Z}_{24}^{\dag}) \langle\tilde{\eta}_{k2}^{\dag}\tilde{\eta}_{-k1}^{\dag}\rangle_{t}
        + (\tilde{Z}^{\dag}_{31}\tilde{Z}_{22}-\tilde{Z}_{32}\tilde{Z}_{34}^{\dag}) \langle\tilde{\eta}_{-k1}\tilde{\eta}_{k2}\rangle_{t} \\
      & \quad\quad\quad
        + (\tilde{Z}^{\dag}_{21}\tilde{Z}_{24}-\tilde{Z}_{34}\tilde{Z}_{24}^{\dag}) \langle\tilde{\eta}_{k2}^{\dag}\tilde{\eta}_{-k2}^{\dag}\rangle_{t}
        + (\tilde{Z}^{\dag}_{41}\tilde{Z}_{22}-\tilde{Z}_{32}\tilde{Z}_{44}^{\dag}) \langle\tilde{\eta}_{-k2}\tilde{\eta}_{k2}\rangle_{t} \\
      & \quad\quad\quad
        + (\tilde{Z}^{\dag}_{31}\tilde{Z}_{24}-\tilde{Z}_{34}\tilde{Z}_{34}^{\dag}) \langle\tilde{\eta}_{-k1}\tilde{\eta}_{-k2}^{\dag}\rangle_{t}
        + (\tilde{Z}^{\dag}_{41}\tilde{Z}_{23}-\tilde{Z}_{33}\tilde{Z}_{44}^{\dag}) \langle\tilde{\eta}_{-k2}\tilde{\eta}_{-k1}^{\dag}\rangle_{t}],
  \end{split}
\end{equation}
and
\begin{equation}\label{eq:steady.value}
  \begin{split}
    C_{mn}(\infty) & = \frac{1}{N'}\sum_{k>0}[
          \tilde{Z}^{\dag}_{11}\tilde{Z}_{21}\langle\tilde{\eta}_{k1}^{\dag}\tilde{\eta}_{k1}\rangle_{t}
         +\tilde{Z}_{31}\tilde{Z}_{14}^{\dag}\langle\tilde{\eta}_{k1}\tilde{\eta}_{k1}^{\dag}\rangle_{t}
         +\tilde{Z}^{\dag}_{21}\tilde{Z}_{22}\langle\tilde{\eta}_{k2}^{\dag}\tilde{\eta}_{k2}\rangle_{t}
         +\tilde{Z}_{32}\tilde{Z}_{24}^{\dag}\langle\tilde{\eta}_{k2}\tilde{\eta}_{k2}^{\dag}\rangle_{t} \\
      &  \quad\quad\quad
         +\tilde{Z}^{\dag}_{31}\tilde{Z}_{23}\langle\tilde{\eta}_{-k1}\tilde{\eta}_{-k1}^{\dag}\rangle_{t}
         +\tilde{Z}_{33}\tilde{Z}_{34}^{\dag}\langle\tilde{\eta}_{-k1}^{\dag}\tilde{\eta}_{-k1}\rangle_{t}
         +\tilde{Z}^{\dag}_{41}\tilde{Z}_{24}\langle\tilde{\eta}_{-k2}\tilde{\eta}_{-k2}^{\dag}\rangle_{t}
         +\tilde{Z}_{34}\tilde{Z}_{44}^{\dag}\langle\tilde{\eta}_{-k2}^{\dag}\tilde{\eta}_{-k2}\rangle_{t} ].
  \end{split}
\end{equation}
It is known that the excitation spectra satisfy the inverse symmetry with respect to $k=0$, i.e. $\Lambda_{k1}=\Lambda_{-k1}$ and $\Lambda_{k2}=\Lambda_{-k2}$. We thus have
\begin{eqnarray}
  \langle\tilde{\eta}_{k1}^{\dag}\tilde{\eta}_{k2}\rangle_{t} = \langle\tilde{\eta}_{-k1}^{\dag}\tilde{\eta}_{-k2}\rangle_{t}  &=& (G_{k1,-k1}^{*}G_{k2,-k1}+G_{k1,-k2}^{*}G_{k2,-k2})e^{it(\tilde{\Lambda}_{k1}-\tilde{\Lambda}_{k2})}, \\
  \langle\tilde{\eta}_{k2}^{\dag}\tilde{\eta}_{k1}\rangle_{t} = \langle\tilde{\eta}_{-k2}^{\dag}\tilde{\eta}_{-k1}\rangle_{t} &=& (G_{k1,-k1}G_{k1,-k2}^{*}+G_{k2,-k2}G_{k2,-k2})e^{-it(\tilde{\Lambda}_{k1}-\tilde{\Lambda}_{k2})}, \\
  \langle\tilde{\eta}_{k1}^{\dag}\tilde{\eta}_{-k1}^{\dag}\rangle_{t} = \langle\tilde{\eta}_{-k1}\tilde{\eta}_{k1}\rangle_{t}^{*} &=& [G_{k1,-k1}^{*}+(G_{k1,-k1}G_{k2,-k2}-G_{k1,-k2}G_{k2,-k1})^{*}G_{k2,-k2}]e^{2it\tilde{\Lambda}_{k1}}, \\
  \langle\tilde{\eta}_{k1}^{\dag}\tilde{\eta}_{-k2}^{\dag}\rangle_{t} = \langle\tilde{\eta}_{-k2}\tilde{\eta}_{k1}\rangle_{t}^{*} &=& [G_{k1,-k2}^{*}-(G_{k1,-k1}G_{k2,-k2}-G_{k1,-k2}G_{k2,-k1})^{*}G_{k2,-k1}]e^{it(\tilde{\Lambda}_{k1}+\tilde{\Lambda}_{k2})}, \\
  \langle\tilde{\eta}_{k2}^{\dag}\tilde{\eta}_{-k1}^{\dag}\rangle_{t} = \langle\tilde{\eta}_{-k1}\tilde{\eta}_{k2}\rangle_{t}^{*} &=& [G_{k2,-k1}^{*}-(G_{k1,-k1}G_{k2,-k2}-G_{k1,-k2}G_{k2,-k1})^{*}G_{k1,-k2}]e^{it(\tilde{\Lambda}_{k1}+\tilde{\Lambda}_{k2})}, \\
  \langle\tilde{\eta}_{k2}^{\dag}\tilde{\eta}_{-k2}^{\dag}\rangle_{t} = \langle\tilde{\eta}_{-k2}\tilde{\eta}_{k2}\rangle_{t}^{*} &=& [G_{k2,-k2}^{*}+(G_{k1,-k1}G_{k2,-k2}-G_{k1,-k2}G_{k2,-k1})^{*}G_{k1,-k1}]e^{2it\tilde{\Lambda}_{k2}}.
\end{eqnarray}
To make the expression of the formula more versatile, we adopt the new symbol $\mathbb{Z}$ to represent the coefficients $\mathbb{Z}_{1} = (\tilde{Z}^{\dag}_{11}\tilde{Z}_{22}-\tilde{Z}_{32}\tilde{Z}_{14}^{\dag}), \cdots$, and use $\mathbb{G}$ to represent the coefficients $\mathbb{G}_{1} = (G_{k1,-k1}^{*}G_{k2,-k1}+G_{k1,-k2}^{*}G_{k2,-k2}), \cdots$. Then we can express the Eq.~(\ref{eq:deltaC.1}) as
\begin{equation}
  \begin{split}
    \delta C_{mn}(t) & = \frac{1}{N}\sum_{k>0}(\mathbb{Z}_{1}+\mathbb{Z}_{12})\mathbb{G}_{1}e^{it(\tilde{\Lambda}_{k1}-\tilde{\Lambda}_{k2})} + (\mathbb{Z}_{2}+\mathbb{Z}_{11})\mathbb{G}_{1}^{*}e^{-it(\tilde{\Lambda}_{k1}-\tilde{\Lambda}_{k2})} \\
      & \quad\quad\quad
        + \mathbb{Z}_{3}\mathbb{G}_{2}e^{2it\tilde{\Lambda}_{k1}} + \mathbb{Z}_{4}\mathbb{G}_{2}^{*}e^{-2it\tilde{\Lambda}_{k1}} + \mathbb{Z}_{9}\mathbb{G}_{5}e^{2it\tilde{\Lambda}_{k2}} + \mathbb{Z}_{10}\mathbb{G}_{5}^{*}e^{-2it\tilde{\Lambda}_{k2}} \\
      & \quad\quad\quad
        + (\mathbb{Z}_{5}\mathbb{G}_{3}+\mathbb{Z}_{7}\mathbb{G}_{4})e^{it(\tilde{\Lambda}_{k1}+\tilde{\Lambda}_{k2})}
      +(\mathbb{Z}_{6}\mathbb{G}_{3}^{*}+\mathbb{Z}_{8}\mathbb{G}_{4}^{*})e^{-it(\tilde{\Lambda}_{k1}+\tilde{\Lambda}_{k2})}.
  \end{split}
\end{equation}

\begin{figure}[t]
  \centering
  \includegraphics[width=0.7\linewidth]{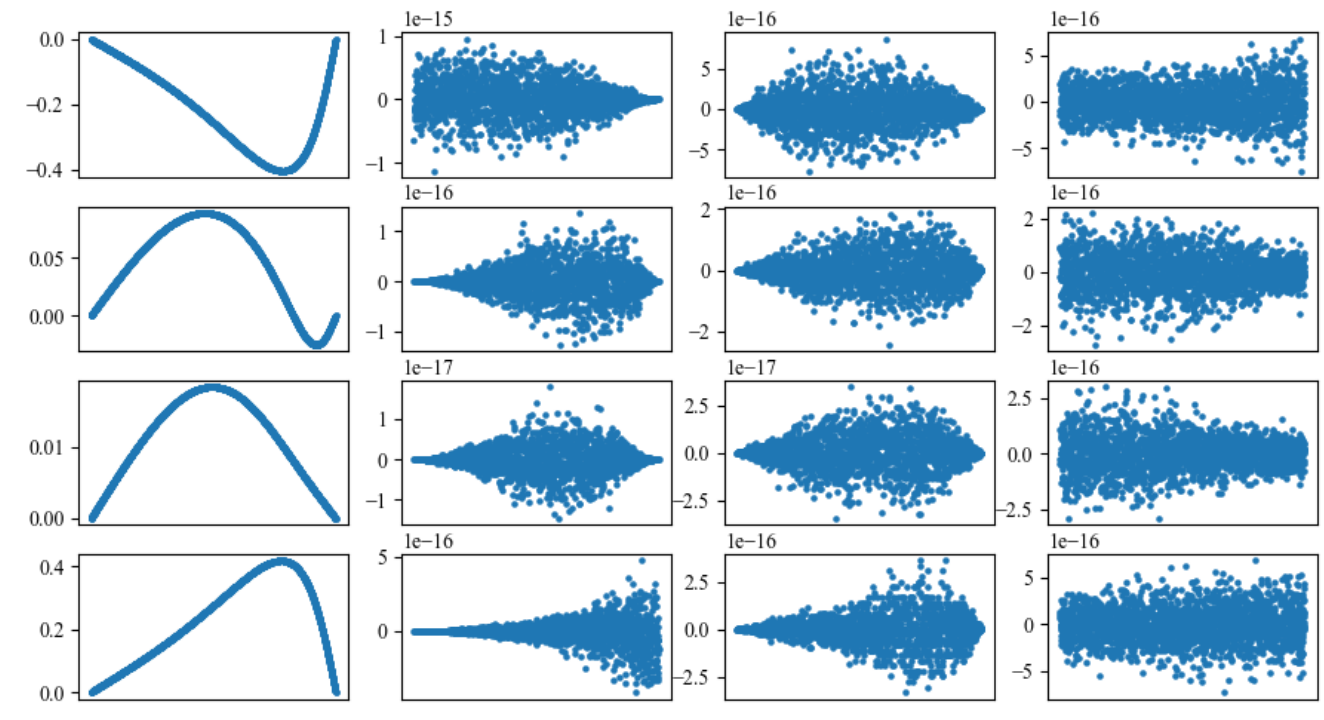}\\
  \caption{Numerical results of coefficients. The results in the first row correspond to $\mathrm{Re}[(\mathbb{Z}_{1}+\mathbb{Z}_{12})\mathbb{G}_{1}+(\mathbb{Z}_{2}+\mathbb{Z}_{11})\mathbb{G}_{1}^{*}]$, $\mathrm{Im}[(\mathbb{Z}_{1}+\mathbb{Z}_{12})\mathbb{G}_{1}+(\mathbb{Z}_{2}+\mathbb{Z}_{11})\mathbb{G}_{1}^{*}]$, $\mathrm{Re}[(\mathbb{Z}_{1}+\mathbb{Z}_{12})\mathbb{G}_{1}-(\mathbb{Z}_{2}+\mathbb{Z}_{11})\mathbb{G}_{1}^{*}]$, and $\mathrm{Im}[(\mathbb{Z}_{1}+\mathbb{Z}_{12})\mathbb{G}_{1}-(\mathbb{Z}_{2}+\mathbb{Z}_{11})\mathbb{G}_{1}^{*}]$, respectively. The results in the second row correspond to $\mathrm{Re}[\mathbb{Z}_{3}\mathbb{G}_{2}+\mathbb{Z}_{4}\mathbb{G}_{2}^{*}]$, $\mathrm{Im}[\mathbb{Z}_{3}\mathbb{G}_{2}+\mathbb{Z}_{4}\mathbb{G}_{2}^{*}]$, $\mathrm{Re}[\mathbb{Z}_{3}\mathbb{G}_{2}-\mathbb{Z}_{4}\mathbb{G}_{2}^{*}]$, and $\mathrm{Im}[\mathbb{Z}_{3}\mathbb{G}_{2}-\mathbb{Z}_{4}\mathbb{G}_{2}^{*}]$, respectively. The results in the third row correspond to $\mathrm{Re}[\mathbb{Z}_{9}\mathbb{G}_{5}+\mathbb{Z}_{10}\mathbb{G}_{5}^{*}]$, $\mathrm{Im}[\mathbb{Z}_{9}\mathbb{G}_{5}+\mathbb{Z}_{10}\mathbb{G}_{5}^{*}]$, $\mathrm{Re}[\mathbb{Z}_{9}\mathbb{G}_{5}-\mathbb{Z}_{10}\mathbb{G}_{5}^{*}]$, and $\mathrm{Im}[\mathbb{Z}_{9}\mathbb{G}_{5}-\mathbb{Z}_{10}\mathbb{G}_{5}^{*}]$, respectively. The results in the fourth row correspond to $\mathrm{Re}[\mathbb{Z}_{5}\mathbb{G}_{3}+\mathbb{Z}_{7}\mathbb{G}_{4}+\mathbb{Z}_{6}\mathbb{G}_{3}^{*}+\mathbb{Z}_{8}\mathbb{G}_{4}^{*}]$, $\mathrm{Im}[\mathbb{Z}_{5}\mathbb{G}_{3}+\mathbb{Z}_{7}\mathbb{G}_{4}+\mathbb{Z}_{6}\mathbb{G}_{3}^{*}+\mathbb{Z}_{8}\mathbb{G}_{4}^{*}]$, $\mathrm{Re}[\mathbb{Z}_{5}\mathbb{G}_{3}+\mathbb{Z}_{7}\mathbb{G}_{4}-\mathbb{Z}_{6}\mathbb{G}_{3}^{*}+\mathbb{Z}_{8}\mathbb{G}_{4}^{*}]$, and $\mathrm{Im}[\mathbb{Z}_{5}\mathbb{G}_{3}+\mathbb{Z}_{7}\mathbb{G}_{4}-\mathbb{Z}_{6}\mathbb{G}_{3}^{*}+\mathbb{Z}_{8}\mathbb{G}_{4}^{*}]$, respectively. }\label{fig:coeffi-1}
\end{figure}

Though numerical calculations (see Fig.~\ref{fig:coeffi-1}), we confirm that
\begin{eqnarray}
  (\mathbb{Z}_{1}+\mathbb{Z}_{12})\mathbb{G}_{1}+(\mathbb{Z}_{2}+\mathbb{Z}_{11})\mathbb{G}_{1}^{*} &=& \mathrm{Re}[(\mathbb{Z}_{1}+\mathbb{Z}_{12})\mathbb{G}_{1}+(\mathbb{Z}_{2}+\mathbb{Z}_{11})\mathbb{G}_{1}^{*}], \\
  (\mathbb{Z}_{1}+\mathbb{Z}_{12})\mathbb{G}_{1}-(\mathbb{Z}_{2}+\mathbb{Z}_{11})\mathbb{G}_{1}^{*} &=& 0, \\
  \mathbb{Z}_{3}\mathbb{G}_{2}+\mathbb{Z}_{4}\mathbb{G}_{2}^{*} &=& \mathrm{Re}[\mathbb{Z}_{3}\mathbb{G}_{2}+\mathbb{Z}_{4}\mathbb{G}_{2}^{*}], \\
  \mathbb{Z}_{3}\mathbb{G}_{2}-\mathbb{Z}_{4}\mathbb{G}_{2}^{*} &=& 0, \\
  \mathbb{Z}_{9}\mathbb{G}_{5}+\mathbb{Z}_{10}\mathbb{G}_{5}^{*} &=& \mathrm{Re}[\mathbb{Z}_{9}\mathbb{G}_{5}+\mathbb{Z}_{10}\mathbb{G}_{5}^{*}], \\
  \mathbb{Z}_{9}\mathbb{G}_{5}-\mathbb{Z}_{10}\mathbb{G}_{5}^{*} &=& 0, \\
  \mathbb{Z}_{5}\mathbb{G}_{3}+\mathbb{Z}_{7}\mathbb{G}_{4}+\mathbb{Z}_{6}\mathbb{G}_{3}^{*}+\mathbb{Z}_{8}\mathbb{G}_{4}^{*} &=& \mathrm{Re}[\mathbb{Z}_{5}\mathbb{G}_{3}+\mathbb{Z}_{7}\mathbb{G}_{4}+\mathbb{Z}_{6}\mathbb{G}_{3}^{*}+\mathbb{Z}_{8}\mathbb{G}_{4}^{*}], \\
    \mathbb{Z}_{5}\mathbb{G}_{3}+\mathbb{Z}_{7}\mathbb{G}_{4}-\mathbb{Z}_{6}\mathbb{G}_{3}^{*}+\mathbb{Z}_{8}\mathbb{G}_{4}^{*} &=& 0, \\
\end{eqnarray}
As a result, we obtain the expression of $\delta C_{mn}(t)$ by
\begin{equation}
  \delta C_{mn}(t) = \int_{0}^{\pi}\frac{dk}{2\pi}[f_{1}(k)\cos{[(\tilde{\Lambda}_{k1}-\tilde{\Lambda}_{k2})t]}+f_{2}(k)\cos{2\tilde{\Lambda}_{k1}t}
  +f_{3}(k)\cos{2\tilde{\Lambda}_{k2}t}+f_{4}(k)\cos{[(\tilde{\Lambda}_{k1}+\tilde{\Lambda}_{k2})t]}],
\end{equation}
where
\begin{eqnarray}
  f_{1}(k) &=& \mathrm{Re}[(\mathbb{Z}_{1}+\mathbb{Z}_{12})\mathbb{G}_{1}+(\mathbb{Z}_{2}+\mathbb{Z}_{11})\mathbb{G}_{1}^{*}], \\
  f_{2}(k) &=& \mathrm{Re}[\mathbb{Z}_{3}\mathbb{G}_{2}+\mathbb{Z}_{4}\mathbb{G}_{2}^{*}], \\
  f_{3}(k) &=& \mathrm{Re}[\mathbb{Z}_{9}\mathbb{G}_{5}+\mathbb{Z}_{10}\mathbb{G}_{5}^{*}], \\
  f_{4}(k) &=& \mathrm{Re}[\mathbb{Z}_{5}\mathbb{G}_{3}+\mathbb{Z}_{7}\mathbb{G}_{4}+\mathbb{Z}_{6}\mathbb{G}_{3}^{*}+\mathbb{Z}_{8}\mathbb{G}_{4}^{*}].
\end{eqnarray}

Similarly, for case 2, we have
\begin{equation}
  C_{mn}(t) = \frac{1}{N'}\sum_{k>0}[\langle\psi_{k}(t)|c_{k2}^{\dag}c_{k1}|\psi_{k}(t)\rangle e^{ik}+\langle\psi_{k}(t)|c_{-k2}^{\dag}c_{-k1}|\psi_{k}(t)\rangle e^{-ik}].
\end{equation}
By considering $c_{k2}^{\dag}c_{k1}=(c_{k1}^{\dag}c_{k2})^{\dag}$ and $c_{-k2}^{\dag}c_{-k1}=(c_{-k1}^{\dag}c_{-k2})^{\dag}$, we can immediately obtain $\delta C_{mn}(t)$ by
\begin{equation}
  \delta C_{mn}(t) = \int_{0}^{\pi}\frac{dk}{2\pi}[f_{1}(k)\cos{[(\tilde{\Lambda}_{k1}-\tilde{\Lambda}_{k2})t]}+f_{2}(k)\cos{2\tilde{\Lambda}_{k1}t}
  +f_{3}(k)\cos{2\tilde{\Lambda}_{k2}t}+f_{4}(k)\cos{[(\tilde{\Lambda}_{k1}+\tilde{\Lambda}_{k2})t]}]\cos{k}.
\end{equation}

\end{widetext}

\bibliography{non-equilibrium}

\end{document}